\documentclass[12pt]{article}

\usepackage{psfrag}
\usepackage[small]{caption}
\usepackage{color}
\definecolor{darkblue}{rgb}{0.1,0.1,.7}
\usepackage[dvips, colorlinks, linkcolor=darkblue, citecolor=darkblue, urlcolor=black, linktocpage]{hyperref}
\usepackage{epsfig,amsmath,amssymb,verbatim,mathrsfs}
\usepackage[square, comma, sort&compress,numbers]{natbib}
\usepackage{dsdshorthand}
\usepackage{feynmp}
\usepackage{subfigure}
\usepackage{simplewick}

\numberwithin{equation}{section}
\newcommand\cN{\mathcal{N}}
\newcommand\cS{\mathcal{S}}
\newcommand\cJ{\mathcal{J}}
\newcommand\bT{\mathbb{T}}
\newcommand\nn{\nonumber}

\newcommand\cR{\mathcal{R}}
\newcommand\lrptl{\raise .8ex\hbox{$^\leftrightarrow$} \hspace{-9pt} \partial}
\newcommand\lptl{\raise .8ex\hbox{$^\leftarrow$} \hspace{-9pt} \partial}
\newcommand\rptl{\raise .8ex\hbox{$^\rightarrow$} \hspace{-9pt} \partial}
\newcommand\twobytwo[4]{\left(\begin{array}{cc}#1 & #2\\ #3 & #4\end{array}\right)}
\newcommand\Vol{\mathop{\mathrm{Vol}}}
\newcommand\jbar{{\bar{\jmath}}}

\title{{\Large {\bf Projectors, Shadows, and Conformal Blocks}}}
\author{
David Simmons-Duffin
\\ \\
{\it \normalsize Jefferson Physical Laboratory, Harvard University}\\
{\it \normalsize Cambridge, MA 02138 USA}\\
}

\begin{document}

\begin{titlepage}

\noindent

\vspace{1cm}

\maketitle
\thispagestyle{empty}

\begin{abstract}
We introduce a method for computing conformal blocks of operators in arbitrary Lorentz representations in any spacetime dimension, making it possible to apply bootstrap techniques to operators with spin.  The key idea is to implement the ``shadow formalism" of Ferrara, Gatto, Grillo, and Parisi in a setting where conformal invariance is manifest.  Conformal blocks in $d$-dimensions can be expressed as integrals over the projective null-cone in the ``embedding space" $\R^{d+1,1}$.  Taking care with their analytic structure, these integrals can be evaluated in great generality, reducing the computation of conformal blocks to a bookkeeping exercise.  To facilitate calculations in four-dimensional CFTs, we introduce techniques for writing down conformally-invariant correlators using auxiliary twistor variables, and demonstrate their use in some simple examples.
\end{abstract}

\end{titlepage}

\setcounter{page}{1}

\setcounter{tocdepth}{2}
\tableofcontents

\vfill\eject


\newpage

\section{Introduction}
\label{sec:intro}

Conformal Field Theories (CFTs) are building blocks for physical theories.  They describe universality classes of myriad quantum and statistical systems in particle and condensed matter physics.  Via the AdS/CFT correspondence \cite{Maldacena:1997re,Gubser:1998bc,Witten:1998qj}, they encode theories of quantum gravity.  Clarifying their structure elucidates all of these diverse fields. 

The dream of the conformal bootstrap program is to classify CFTs using symmetries and consistency conditions alone \cite{Mack:1969rr,Ferrara:1972xe,Ferrara:1972ay,Ferrara:1972uq,Ferrara:1973vz,Polyakov:1974gs}.  The key principles (the Operator Product Expansion (OPE), crossing symmetry, and unitarity) have been understood for decades, and applied with success in 2d, where the infinite-dimensional Virasoro symmetry provides another powerful tool \cite{Polyakov:1984yq}.  However, for CFTs in $d>2$, we are only now learning how to obtain concrete results \cite{Rattazzi:2008pe,Rychkov:2009ij,Caracciolo:2009bx,Poland:2010wg,Rattazzi:2010gj,Rattazzi:2010yc,Vichi:2011ux,Poland:2011ey,Rychkov:2011et,ElShowk:2012ht}.
   
Current bootstrap methods rely crucially on expressions for conformal blocks, which encode the contribution of a primary operator $\cO$ to a four-point function of primary operators $\<\f_1\f_2\f_3\f_4\>$.  We will introduce a method for computing conformal blocks of operators in arbitrary Lorentz representations, making it possible to study the full implications of crossing symmetry and unitarity in CFTs. 
Compact expressions for scalar conformal blocks in two and four dimensions \cite{DO1,DO2} were important in the initial discovery of universal bounds on operator dimensions and OPE coefficients \cite{Rattazzi:2008pe,Rychkov:2009ij,Caracciolo:2009bx}.  Later, they were essential for improving these methods and deriving bounds with concrete phenomenological implications \cite{Rattazzi:2010yc,Vichi:2011ux,Poland:2011ey}.  Computations of scalar superconformal blocks allowed for similar bounds in 4d superconformal theories \cite{Poland:2010wg,Vichi:2011ux,Poland:2011ey}.  More recently, an improved understanding of scalar conformal blocks in 3d led to novel determinations of operator dimensions in the 3d Ising model \cite{ElShowk:2012ht}, with precision comparable to the best perturbative calculations and Monte-Carlo simulations.

All of these results come from studying crossing symmetry and unitarity for a four-point function of scalars.  But this is a small subset of the full consistency conditions of a CFT.  Why not study correlators of more general operators, not just scalars?  For example, applying bootstrap methods to four-point functions of currents $\<J^{\mu}J^{\nu}J^{\rho}J^{\sigma}\>$ or the stress-tensor $\<T^{\mu\nu}T^{\rho\sigma}T^{\kappa\lambda}T^{\alpha\beta}\>$ might lead to universal bounds on symmetry representations and central charges, perhaps shedding light on the bounds on $a,c$ in \cite{Hofman:2008ar}, or the weak gravity conjecture \cite{ArkaniHamed:2006dz}.

Such investigations would require expressions for conformal blocks of operators with spin.  Unfortunately, these are scarce.  Methods for computing scalar blocks can become intractable in the case of higher spin.  For example, Dolan and Osborn derived scalar blocks by solving an eigenvalue equation for the quadratic-Casimir of the conformal group, which takes the form of a single second-order PDE \cite{DO2}.  But because of the many tensor structures that can enter a four-point function of spin-1 operators, the analogous equation for conformal blocks of spin-1 operators is a system of 43 coupled second-order PDEs.\footnote{Experience has shown that it is sometimes sufficient to compute recursion relations which allow for efficient numerical computation and tabulation of conformal blocks and their derivatives \cite{ElShowk:2012ht}, so completely explicit solutions are not obligatory.  It is possible that one can effectively compute numerical approximations to higher spin conformal blocks, perhaps along the lines of the methods recently developed in \cite{Hogervorst:2013sma,Hogervorst:2013kva,Kos:2013tga}.}

Partial progress on this problem was made recently in \cite{Costa:2011dw,Costa:2011mg}, where the authors leveraged existing results for scalar blocks to write down conformal blocks for traceless symmetric tensors (TSTs) of the Lorentz group.  This is sufficient for bootstrapping 3d CFTs, where TSTs exhaust the list of bosonic Lorentz representations.  However in $d>3$, it is insufficient.

In this work, we present a general method for computing conformal blocks of operators in arbitrary Lorentz representations.  The underlying idea is based on the shadow formalism of Ferrara, Gatto, Grillo, and Parisi \cite{Ferrara:1972xe,Ferrara:1972ay,Ferrara:1972uq,Ferrara:1973vz}.  Given an operator $\cO(x)$ with dimension $\De$ in a $d$-dimensional CFT, they define a nonlocal ``shadow operator" $\tl\cO(x)$ with dimension $\tl\De=d-\De$.  The integral
\begin{align}
\label{eq:ferraraparisicompleteness}
\int d^dx\, \cO(x)|0\>\<0|\tl\cO(x)
\end{align}
is then dimensionless and invariant under conformal transformations.  When inserted between pairs of operators, it {\it almost} does the job of projecting onto the contribution of $\cO$ to a four-point function --- the conformal block $g_{\cO}$,
\begin{align}
\int d^dx\<\f_1(x_1)\f_2(x_2) \cO(x)\>\<\tl\cO(x)\f_3(x_3)\f_4(x_4)\>
&= g_\cO(x_i) + \textrm{``shadow block"}.
\end{align}
The extra ``shadow block" is distinguished from $g_\cO(x_i)$ by its behavior as $x_{12}\to0$, and needs to be subtracted off.

A challenge in applying this procedure to operators with spin is defining conformally-invariant projectors analogous to~(\ref{eq:ferraraparisicompleteness}).  For this, it is extremely useful to use the embedding-space formalism \cite{Dirac:1936fq,Mack:1969rr,Boulware:1970ty,Ferrara:1973eg,Weinberg:2010fx}, which makes conformal invariance manifest by linearizing the action of the conformal group.  In section \ref{sec:constructingconformalblocks}, we introduce the shadow formalism in this context, using scalar conformal blocks as an example but casting shadows into a form which readily generalizes to higher spin.  We show how~(\ref{eq:ferraraparisicompleteness}) can be understood as a manifestly conformally-invariant integral over the projective null-cone in $\R^{d+1,1}$, called a ``conformal integral."  The utility of writing conformally-invariant integrals in projective space has already been recognized to some extent in loop calculations for amplitudes \cite{Hodges:2010kq,Mason:2010pg,Paulos:2012nu}.  In this work, it will be crucial both for ensuring conformal invariance and simplifying calculations.  Also in section~\ref{sec:constructingconformalblocks}, we give a simple way to disentangle the conformal block $g_\cO(x_i)$ from its shadow by considering the action of a monodromy $x_{1,2}\to e^{2\pi i}x_{1,2}$.

In section~\ref{sec:conformalintegrals}, we compute all conformal integrals which arise in conformal block computations, and clarify their properties under monodromy.  Using the embedding space, integrals with nontrivial Lorentz indices are no more difficult than scalar integrals, and the results of this section apply equally well to scalar and higher-spin blocks.  In even spacetime dimensions, the expressions are sums of products of elementary hypergeometric functions. In section~\ref{sec:higherspinconformalblocks}, we explain the strategy for combining these results to compute higher spin conformal blocks.  As an example, we write down conformally-invariant projectors for tensor operators, and compute the conformal block for the exchange of an antisymmetric tensor in a four-point function of scalars and  spin-1 operators.

In section~\ref{sec:twistormethods}, we specialize to the case of CFTs in four-dimensions.  We develop a formalism for studying correlators of operators in arbitrary Lorentz representations using auxiliary twistor variables.  Within this formalism, we define projectors and shadows for multi-twistor operators and then demonstrate their use for computing conformal blocks in a few simple examples.  We conclude in section~\ref{sec:conclusion}.

\section{Constructing Conformal Blocks}
\label{sec:constructingconformalblocks}

\subsection{Defining Properties}
\label{sec:conformalblocks}

A conformal block encodes the contribution of a single irreducible conformal multiplet (a primary operator and its descendants) to a four-point function of primary operators.  Consider, for example, a four-point function of primary scalars $\f_i$ with dimensions $\De_i$.  We can expand it as a sum over conformal multiplets by inserting a complete set of states,
\begin{align}
\<\f_1(x_1)\f_2(x_2)\f_3(x_3)\f_4(x_4)\> &= \sum_{\cO\in \f_1\x\f_2}\,\sum_{\a=\cO,P\cO,\dots}\<0|\f_3(x_3)\f_4(x_4)|\a\>\<\a|\f_1(x_1)\f_2(x_2)|0\>.
\end{align}
Here, $\cO\in \f_1\x\f_2$ runs over {\it primary} operators $\cO$ appearing in the OPE of $\f_1\x\f_2$, and $\a$ runs over $\cO$ and its (normalized) descendants, considered as states in radial quantization on a sphere separating $x_1,x_2$ from $x_3,x_4$.

For fixed $\cO$, the quantities $\<\a|\f_i(x_i)\f_j(x_j)|0\>$ are proportional to the three-point function coefficient $\l_{\f_i\f_j\cO}$.  Stripping these off, we are left with a purely kinematical quantity called a conformal partial wave,
\begin{align}
W_\cO(x_i) &= \frac 1 {\l_{\f_1\f_2\cO}\l_{\f_3\f_4\cO}}\sum_{\a=\cO,P\cO,\dots}\<0|\f_3(x_3)\f_4(x_4)|\a\>\<\a|\f_1(x_1)\f_2(x_2)|0\>.
\end{align}
The conformal block $g_\cO(u,v)$ is defined in terms of $W_\cO(x_i)$ by additionally removing factors of $x_{ij}^2$ to obtain a dimensionless quantity,
\begin{align}
\label{eq:conformalblockdefinition}
W_\cO(x_i)
&=
\p{\frac{x_{14}^2}{x_{13}^2}}^{\frac{\De_{34}}{2}}\p{\frac{x_{24}^2}{x_{14}^2}}^{\frac{\De_{12}}2}\frac{g_{\cO}(u,v)}{x_{12}^{\De_1+\De_2}x_{34}^{\De_3+\De_4}},
\end{align}
where $\De_{ij}\equiv \De_i-\De_j$, and $u=\frac{x_{12}^2 x_{34}^2}{x_{13}^2 x_{24}^2}$, $v=\frac{x_{14}^2 x_{23}^2}{x_{13}^2 x_{24}^2}$ are conformally-invariant cross-ratios.

The form of $g_{\cO}(u,v)$ is completely fixed by conformal symmetry, and depends only on the representations of $\cO$ and the $\f_i$ under the conformal group (i.e. their dimensions and spins).
One way to see why is to note that $g_{\cO}(u,v)$ possesses the following three properties:
\begin{enumerate}
\item It is invariant under conformal transformations.

\item It is an eigenvector of the quadratic Casimir of the conformal group acting on $x_1,x_2$.  Specifically, let $L_A$, with $A=1,\dots,(d+1)(d+2)/2$ indexing the adjoint of $\SO(d+1,1)$, be generators of conformal transformations, and denote the associated differential operators acting on $\f_i(x_i)$ by $\cL_{iA}$: $\cL_{iA}\f_i(x_i)=[\f_i(x_i),L_A]$.  Each descendant $|\a\>$ is an eigenvector of $L^AL_A$ with the same eigenvalue $C_\cO=\De(d-\De)+C_{L}$, where $\De$ is the dimension of $\cO$, and $C_{L}$ is the Casimir of the Lorentz representation of $\cO$.  Thus,
\begin{align}
&(\cL_{1A}+\cL_{2A})(\cL_1^{A}+\cL_2^{A}) W_\cO(x_i)\nn\\
 &= \frac 1 {\l_{\f_1\f_2\cO}\l_{\f_3\f_4\cO}}\sum_\a\<0|\f_3(x_3)\f_4(x_4) |\a\>\<\a| [[\f_1(x_1)\f_2(x_2),L_A],L^A]|0\>\nn\\
&= C_{\cO} W_\cO.
\label{eq:eigenvalueEq}
\end{align}
Eqs.~(\ref{eq:eigenvalueEq}) and~(\ref{eq:conformalblockdefinition}) then imply an eigenvalue equation for $g_\cO(u,v)$.

\item The behavior of $g_\cO(u,v)$ as $x_{12}\to0$ 
is dictated by the primary term $\cO\in \f\x\f$ in the OPE.
More explicitly, if $\cO$ is a spin-$\ell$ operator, we have
\begin{align}
\f_1(x_1)\f_2(x_2)=&\l_{\f_1\f_2\cO}x_{12}^{\De-\De_1-\De_2-\ell}x_{12\mu_1}\dots x_{12\mu_\ell}\cO^{\mu_1\dots\mu_\ell}(x_2)\nn\\
&+\textrm{descendants} + \textrm{other multiplets}.
\end{align}
Descendants of $\cO$ come with higher powers of $x_{12}$ in the OPE, and other multiplets do not contribute to $g_\cO$.  Hence the small $x_{12}$ limit of our conformal block comes from the leading term above,
\begin{align}
g_\cO(u,v) &\sim x_{12}^{\De-\ell}x_{12\mu_1}\dots x_{12\mu_\ell} \<\cO^{\mu_1\dots\mu_\ell}(x_2)\f_3(x_3)\f_4(x_4)\>.
\label{eq:limitingValue}
\end{align}

\end{enumerate}

Together, these properties determine $g_\cO(u,v)$.  This is demonstrated for example in \cite{DO2} where Dolan and Osborn explicitly solve~(\ref{eq:eigenvalueEq}) subject to~(\ref{eq:limitingValue}).  In even dimensions, their solution takes a simple form in terms of hypergeometric functions.  For instance when $d=4$,\footnote{Our normalization of $g_\cO(u,v)$ differs by a factor of $2^\ell$ from the one in \cite{DO2}.}
\begin{align}
\label{eq:DOscalarblock}
g_\cO(u,v) &= (-1)^\ell\frac{z\bar z}{z-\bar z}(k_{\De+\ell}(z)k_{\De-\ell-2}(\bar z)-z\leftrightarrow \bar z)\nn\\
k_\b(x) &\equiv x^{\b/2}{}_2F_1\p{\frac{\b-\De_{12}}{2},\frac{\b+\De_{34}}{2},\b;x},
\end{align}
where $\De,\ell$ are the dimension and spin of $\cO$, respectively, and $z$ and $\bar z$ are defined in terms of the cross ratios $u$ and $v$ by
\begin{align}
u&=z\bar z,\qquad v=(1-z)(1-\bar z).
\end{align}

In more general situations, the conformal Casimir equation becomes a complicated system of coupled PDEs that can be difficult to solve.  Instead of solving it directly, our approach will be to write down expressions that manifestly satisfy properties 1, 2, and 3, and then compute them.  This method, essentially the shadow formalism \cite{Ferrara:1972xe,Ferrara:1972ay,Ferrara:1972uq,Ferrara:1973vz}, was used in Dolan and Osborn's original derivation of~(\ref{eq:DOscalarblock}) \cite{DO1}.  Our contribution will be to clarify and generalize this approach, providing a unified way to ensure each of the above properties holds, along with a toolkit for performing the resulting calculations.  To this end, let us address each property in turn.

\subsection{The Embedding Space}

The constraints of conformal invariance are most transparent in the embedding space \cite{Dirac:1936fq,Mack:1969rr,Boulware:1970ty,Ferrara:1973eg,Weinberg:2010fx}.  Consider a Euclidean CFT in $d$-dimensions, with conformal group $\SO(d+1,1)$ acting nonlinearly on spacetime $\R^d$.  The key idea, originally due to Dirac, is that this nonlinear action is induced from the much simpler linear action of $\SO(d+1,1)$ on the ``embedding space" $\R^{d+1,1}$.  To see how, 
choose coordinates $X^m=(X^+,X^-,X^\mu)$ on $\R^{d+1,1}$, with the inner product
\begin{align}
\label{eq:embeddinginnerproduct}
X\.X &= \eta_{mn}X^m X^n= -X^+X^-+X_\mu X^\mu.
\end{align}
The condition $X^2=0$ defines an $\SO(d+1,1)$-invariant subspace of dimension $d+1$ --- the null-cone.  We obtain $d$-dimensional Euclidean space by projectivizing: quotienting the null-cone by the rescaling $X\sim \l X$, $\l\in \R$.  Because projectivizing respects Lorentz rotations of the embedding space, the projective null-cone naturally inherits an action of $\SO(d+1,1)$.

We can identify the projective null-cone with $\R^d$ by ``gauge-fixing" this rescaling.  For example, imposing the gauge condition $X^+=1$, null vectors take the form $X=(1,x^2,x^\mu)$, for $x^\mu\in \R^d$.\footnote{This gauge condition fails for precisely one null direction, $X=(0,1,0)$ representing the point at infinity.}  This gauge slice is called the {\it Poincar\'e section}.
Beginning with some point $X=(1,x^2,x^\mu)$, a transformation $h\in\SO(d+1,1)$ takes $X$ to $hX$ by matrix multiplication.  To get back to the Poincar\'e section, we must further rescale $hX \to hX/(hX)^+$.  The combined transformation $X\to hX/(hX)^+$ is precisely the nonlinear action of the conformal group on $\R^d$.  Note that on the Poincar\'e section, we have
\begin{align}
-2X\.Y=(x-y)^2.
\end{align}

Primary operators on $\R^d$ can be lifted to homogeneous, conformally-covariant fields on the null-cone.  For example, given a primary scalar $\f(x)$ with dimension $\De$, one can define a scalar on the entire null-cone by
\begin{align}
\Phi(X) &\equiv (X^+)^{-\De }\f(X^\mu/X^+).
\end{align}
The field $\Phi(X)$ then transforms simply under conformal transformations $\Phi(X)\to \Phi(hX)$.  Conformal invariance means that correlators of $\Phi(X)$ are invariant under linear $\SO(d+1,d)$ rotations.

The dimension of $\f$ is reflected in the degree of $\Phi$,
\begin{align}
\Phi(\l X) &= \l^{-\De}\Phi(X).
\end{align}
This homogeneity condition must be respected by any correlator involving $\Phi(X)$.  For example, the two-point function $\<\Phi(X_1)\Phi(X_2)\>$ is fixed by conformal invariance, homogeneity, and the null condition $X_i^2=0$ to have the form
\begin{align}
\<\Phi(X_1)\Phi(X_2)\> &\propto \frac{1}{X_{12}^\De},
&
X_{ij} \equiv -2 X_i\. X_j.
\label{eq:scalar2ptfunction}
\end{align}
The notation $X_{ij}$ is for convenience when comparing to flat-space coordinates on the Poincar\'e section, $X_{ij}\to x_{ij}^2$.  In our conventions, $\Phi$ is canonically normalized when the constant of proportionality in~(\ref{eq:scalar2ptfunction}) is $1$.

 One can additionally lift fields with spin to conformally covariant fields on the null-cone \cite{Costa:2011dw}.  We defer discussion of this machinery until it is needed in section~\ref{sec:spinningcorrelatorsformalism}.  In what follows, we will write simply $\f(X)$ to indicate the lift of $\f(x)$ to the embedding space.

\subsection{Conformal Integrals}

The projective null-cone admits a natural notion of integration that produces new conformal invariants from old ones.  Let us start with an obvious $\SO(d+1,1)$-invariant measure on the null-cone, $d^{d+2}X\,\de(X^2)$, where $\de(X^2)$ is a Dirac delta-function.  This measure has degree $d$ in $X$, so only its product with a degree $-d$ function $f(X)$ is well-defined after projectivization.  However the integral
\begin{align}
\int d^{d+2}X\,\de(X^2) f(X)
\end{align}
is formally infinite because of the rescaling invariance $X\to\l X$.

We can obtain a finite result by dividing by the volume of the ``gauge-group."\footnote{Precisely, we quotient by  the connected component of the identity $\GL(1,\R)^+\subset\GL(1,\R)$ and restrict the integral to a single branch of the null cone.} Specifically, let us define 
\begin{align}
\label{eq:conformalintegraldefinition}
\int D^d X f(X) &\equiv \frac{2}{\Vol \GL(1,\R)^+}\int_{X^++X^-> 0} d^{d+2}X\de(X^2) f(X).
\end{align}
Integrals of this form, which we call ``conformal integrals," will play a central role in this work.\footnote{An alternative definition of the conformal integral measure is as a residue $D^d X=\frac 1 {2\pi i}\oint_{S^1} \frac{\w}{X^2}$, where $\w=\frac 1 {(d+1)!}\e_{m_0\dots m_{d+1}}X^{m_0}dX^{m_1}\we\cdots \we dX^{m_{d+1}}$ is an $\SO(d+2)$-invariant volume form on projective space $\P^{d+1}$, and the $S^1$ encircles the locus where $X^2=0$.  The combination $\frac{\w}{X^2}$ has projective weight $d$, so it can be integrated against a section with projective weight $-d$.  The full integration contour we consider has topology $S^1\x S^d$.}
In practice, we can evaluate them by gauge-fixing and supplying the appropriate Faddeev-Popov determinant.  For example, the gauge choice $X^+=1$ reduces~(\ref{eq:conformalintegraldefinition}) to a conventional integral over flat space.  The advantage of the definition~(\ref{eq:conformalintegraldefinition}) is that it makes $\SO(d+1,1)$-invariance manifest.

As an example, let us evaluate a conformal integral which will be important in subsequent computations,
\begin{align}
\label{eq:basicconformalintegral}
I(Y) &= \int D^d X \frac{1}{(-2X\.Y)^d}\qquad (Y^2 < 0).
\end{align}
Note that this is essentially the unique conformal integral depending on a single vector $Y$ and producing a scalar.  The requirement that the integrand have degree $-d$ in $X$, along with the null condition $X^2=0$ fixes the integrand up to a constant.

Since $I(Y)$ is conformally invariant and homogeneous in $Y$, we are free to choose $Y=Y_0=(1,1,0)$ with $Y_0^2=-1$, and recover the full $Y$-dependence at the end from dimensional analysis.  From the definition of the  measure~(\ref{eq:conformalintegraldefinition}), we have
\begin{align}
I(Y_0) &= \frac 2 {\Vol \GL(1,\R)^+}\int_{X^++X^-\geq 0} \frac 1 2 dX^- dX^+ d^d X  \de(-X^+ X^- + X_\mu X^\mu) \frac{1}{(X^++X^-)^d}\nn\\
&= \frac 1 {\Vol \GL(1,\R)^+}\int d^{d}X \int_0^\oo \frac{dX^+}{X^+}\frac{1}{(X^++X_\mu X^\mu/X^+)^d}\nn\\
&= \int d^d X \frac 1 {(1+X_\mu X^\mu)^d}= \frac{\pi^{d/2}\G(d/2)}{\G(d)}.
\end{align}
In the third line, we have made the gauge choice $X^+=1$.  The associated Faddeev-Popov determinant is $1$.  Restoring the factors of $-Y^2$ required by dimensional analysis gives
\begin{align}
I(Y) &= \frac{\pi^{d/2}\G(d/2)}{\G(d)} \frac{1}{(-Y^2)^{d/2}}.
\label{eq:basicconformalintegralanswer}
\end{align}

As a consistency check, let us evaluate this integral by gauge-fixing in a different way. Let us write $X^\pm=X^{-1}\pm X^0$, so that the metric (\ref{eq:embeddinginnerproduct}) takes the form $X^2 = -(X^{-1})^2 + \sum_{i=0}^d X^i X^i$. We now choose the gauge $X^{-1}=1$, so that the gauge-fixed integral is over a sphere $S^d$. Again, the Fadeev-Popov determinant is $1$, so we have
\be
I(Y_0) &=& 2 \int d^{d+1} X \de\p{-1+\sum_{i=0}^d X^i X^i} \left.\frac{1}{(-2X\.Y_0)^d}\right|_{X^{-1}=1} \nn\\
&=& 2 \mathrm{vol}(S^d) \int r^d dr \de(-1+r^2) \frac{1}{2^d} \nn\\
&=& \frac{\mathrm{vol}(S^d)}{2^d} = \frac{\pi^{d/2}\G(d/2)}{\G(d)}.
\ee

Eq.~(\ref{eq:basicconformalintegralanswer}) is sufficient for evaluating numerous conformal integrals.  For instance, products $\prod_i(-2X\.Y_i)^{-a_i}$ can be reduced to the form~(\ref{eq:basicconformalintegral}) using the Feynman/Schwinger parameterization
\begin{align}
\label{eq:schwingerformula}
\frac{1}{\prod_i A_i^{a_i}} &= \frac{\G\p{\sum_i a_i}}{\prod_i \G(a_i)}\int_0^\oo \prod_{i=2}^n d\a_i\,\a_i^{a_i-1}\frac{1}{(A_1+\sum_{i=2}^n \a_i A_i)^{\sum_i a_i}}.
\end{align}
Combining~(\ref{eq:schwingerformula}) and~(\ref{eq:basicconformalintegralanswer}), a three-point integral is given by
\begin{align}
\label{eq:threepointconformalintegral}
\int D^{d} X_0 \frac{1}{X_{10}^a X_{20}^b X_{30}^c}
&=
\frac{\pi^h \G(h-a)\G(h-b)\G(h-c)}{\G(a)\G(b)\G(c)}\frac{1}{X_{12}^{h-c}X_{13}^{h-b}X_{23}^{h-a}},
\end{align}
where $h\equiv d/2$ and $a+b+c=d$ so that the projective measure is well-defined.  Note that the form of this result is fixed by homogeneity in $X_1,X_2,X_3$.

More generally, any conformal integral can be manipulated to a sum of terms of the form
\begin{align}
\int D^d X \frac{X^{m_1}\dots X^{m_n}}{(-2X\.Y)^{d+n}}
&=
\frac{\G(d)}{2^n \G(d+n)}\p{\prod_i \pdr{}{Y_{m_i}}} I(Y)\nn\\
&= \frac{\pi^{d/2}\G(d/2+n)}{\G(d+n)}\frac{Y^{m_1}\cdots Y^{m_n}}{(-Y^2)^{d/2+n}} - \textrm{traces},
\label{eq:conformalintegralwithindices}
\end{align}
where traces are subtracted using the embedding space metric $\eta_{mn}$.  Tracelessness is clear in the integrand because $X^2$ vanishes on the null-cone.  Eqs~(\ref{eq:basicconformalintegralanswer}) and~(\ref{eq:conformalintegralwithindices}) undergird most of the computations in this work.

\subsection{The Conformal Casimir}

Three-point functions of primary operators provide natural eigenvectors of the conformal Casimir.\footnote{Note that the differential operators generating conformal transformations in the embedding space are just the usual generators of $\SO(d+1,1)$ acting on functions on $\R^{d+1,1}$, $\cL_{mn} = X_m\pdr{}{X^n}-X_n\pdr{}{X^m}$.}  Because $\<\f_1\f_2\cO\>$ is conformally covariant, we have
\begin{align}
(\cL_{1A}+\cL_{2A})\<\f_1(X_1)\f_2(X_2)\cO(X_3)\> 
&=
-\cL_{3A}\<\f_1(X_1)\f_2(X_2)\cO(X_3)\>.
\end{align}
Thus, action of the conformal Casimir on $X_1,X_2$ is equivalent to action on $X_3$, which gives simply the eigenvalue  $C_\cO$,
\begin{align}
(\cL_{1A}+\cL_{2A})(\cL^{1A}+\cL^{2A})\<\f_1(X_1)\f_2(X_2)\cO(X_3)\>
&= \cL^{3A}\cL_{3A}\<\f_1(X_1)\f_2(X_2)\cO(X_3)\>\nn\\
&= C_\cO\<\f_1(X_1)\f_2(X_2)\cO(X_3)\>.
\end{align}
This argument is independent of the actual value of $X_3$, so any linear combination of $\<\f_1(X_1)\f_2(X_2)\cO(X_3)\>$ with different values of $X_3$ is also an eigenvector of the conformal Casimir acting on $X_1,X_2$, with the same eigenvalue.  In particular, so is the conformal integral
\begin{align}
\label{eq:eigenvalueform}
\int D^dX_3 \<\f_1(X_1)\f_2(X_2)\cO(X_3)\>f(X_3)
\end{align}
where $f(X)$ is any homogeneous function on the null-cone with degree $\De-d$.

This suggests a natural candidate for the conformal partial wave $W_\cO$,
\begin{align}
\label{eq:candidateconformalblock}
W_\cO(X_i) &\stackrel{?}= 
  \frac{1}{\cN_{\cO}}
\int D^d X D^d Y \<\f_1(X_1)\f_2(X_2)\cO(X)\>\frac{1}{(-2 X\.Y)^{d-\De}}\<\cO(Y)\f_3(X_3)\f_4(X_4)\>,
\end{align}
where $\cN_\cO$ is a constant to be determined.  Note that~(\ref{eq:candidateconformalblock}) has the correct degree in the $X_i$, is manifestly conformally invariant, and is also manifestly an eigenvector of the conformal Casimir acting on $X_1,X_2$ (equivalently $X_3,X_4$) with the correct eigenvalue, since it has the form~(\ref{eq:eigenvalueform}).  The denominator $(-2X\.Y)^{d-\De}$ is the unique choice for which the conformal integrals over $X$ and $Y$ are well-defined.  We will see shortly that~(\ref{eq:candidateconformalblock}) is incorrect, but is a convenient stepping stone to the correct answer.

We can rewrite~(\ref{eq:candidateconformalblock}) in a useful way by introducing the {\it shadow operator},
\begin{align}
\label{eq:shadowdefinition}
\tl \cO(X) &= \int D^d Y\frac{1}{(-2 X\.Y)^{d-\De}}\cO(Y),
\end{align}
which formally has the transformation properties of a primary scalar with dimension $d-\De$.\footnote{Since $\tl\cO$ is nonlocal, this does not contradict unitarity.}  Note that $\tl\cO$ has the same eigenvalue as $\cO$ under the conformal Casimir, since $C_\cO$ is invariant under $\De\to d-\De$. In terms of $\tl\cO$, eq.~(\ref{eq:candidateconformalblock}) reads
\begin{align}
\label{eq:opshadowopintegral}
W_\cO(X_i) &\stackrel{?}= \frac{1}{\cN_\cO} \int D^d X_0 \<\f_1(X_1)\f_2(X_2)\cO(X_0)\>\<\tl\cO(X_0)\f_3(X_3)\f_4(X_4)\>.
\end{align}

For example, in the special case where the $\De_i$ are all equal to $\de$, we have the candidate conformal block
\begin{align}
g_\cO(X_i) &\stackrel{?}= 
 \frac1{\cN_\cO}
 \frac{\pi^{d/2}\G(\De-\frac d 2)\G^2(\frac{d-\De}{2})}{\G(d-\De)\G^2(\frac{\De}2)}
  \int D^d X_0 \frac{X_{12}^{\De/2}X_{34}^{(d-\De)/2}}{X_{10}^{\De/2}X_{20}^{\De/2}X_{03}^{(d-\De)/2}X_{04}^{(d-\De)/2}}
 \equiv F(X_i),
\label{eq:naiveconformalintegral}
\end{align}
where again $X_{ij}\equiv -2 X_i\.X_j$, and we have evaluated $\<\tl\cO(X_0) \f_3(X_3)\f_4(X_4)\>$ using~(\ref{eq:threepointconformalintegral}).

\subsection{Consistency with the OPE}

Eq.~(\ref{eq:naiveconformalintegral}) is a conformally-invariant eigenvector of the conformal Casimir with the correct eigenvalue.  Our final requirement is that it have the correct limiting behavior as $X_{12}\to 0$,  namely $g_{\cO}(X_i)\sim X_{12}^{\De/2}$.   This is indeed the behavior of the {\it integrand} above.  But the full behavior of the {\it integral} $F(X_i)$ is unclear.  The integral over $X_0$ could potentially probe the region near $X_1,X_2$ in ways that introduce new singularities.

In fact this {\it must} happen, since we could have performed the $X$ integral in~(\ref{eq:candidateconformalblock}) first, exchanging $\De\leftrightarrow  d-\De$ in the integrand of~(\ref{eq:naiveconformalintegral}).  Symmetry under $\De\leftrightarrow  d-\De$ implies that $F(X_i)$ must actually compute a linear combination of the conformal block $g_{\cO}$ and its shadow block $g_{\tl\cO}$ (which has the same eigenvalue under the conformal Casimir, but different limiting behavior $g_{\tl\cO}(X_i)\sim X_{12}^{(d-\De)/2}$ as $X_{12}\to 0$).  In other words,
\begin{align}
F(X_i) = g_{\cO}(X_i) + K_\cO g_{\tl\cO}(X_i),
\end{align}
where $K_\cO$ is a constant.

Thus, our final step should be to remove the shadow component $g_{\tl\cO}(X_i)$ from $F(X_i)$.  This procedure can be performed quickly and elegantly in Mellin space \cite{Fitzpatrick:2011hu}, but takes some care in position space.  The approach of \cite{DO1} is to evaluate integrals like~(\ref{eq:naiveconformalintegral}) as a series in conformal cross ratios $u,1-v$, discard terms of the form $u^{(d-\De)/2+n}(1-v)^m$, $m,n\in\Z$, which belong to the shadow block, and re-sum the remaining terms.

Here, we will take a cleaner approach that avoids complicated series expansions and special function identities.  The key observation is that $g_{\cO}$ and $g_{\tl \cO}$ are distinguished by their behavior under monodromy $M:X_{12}\to e^{4\pi i} X_{12}$,\footnote{$M$ can be generated by exponentiating a dilatation operator $e^{2\pi i(\cD_1+\cD_2)}$ acting on $X_1,X_2$.  See appendix~\ref{app:monodromyandOPE} for details.}
\begin{align}
M: g_\cO &\to e^{2\pi i \De}g_{\cO}\\
M: g_{\tl \cO} &\to e^{2\pi i (d-\De)}g_{\tl\cO}.
\end{align}
Isolating $g_\cO$ means projecting $F(X_i)$ onto the correct eigenspace of $M$,
\begin{align}
g_\cO(X_i) &= F(X_i)|_{M=e^{2\pi i \De}}.
\label{eq:conformalblockfinalexpression}
\end{align}
Since $M$ commutes with conformal transformations, so does projection onto its eigenspaces.  Consequently, (\ref{eq:conformalblockfinalexpression}) is still conformally invariant, and still solves the correct Casimir differential equation.  Thus, it satisfies the requirements for a conformal block, and all that remains is to compute it.  We will do so in section~\ref{sec:conformalintegrals}.  We give more detail about how monodromy projection ensures the correct small $x_{12}$ behavior and why the shadow block $g_{\tl\cO}$ appears in appendix~\ref{app:monodromyandOPE}.

\subsection{Projectors and Shadows}
\label{sec:projectorsandshadows}

Our prescription~(\ref{eq:opshadowopintegral}) for computing conformal blocks can be summarized succinctly as the statement that
\begin{align}
\label{eq:projectionoperatorintermsofshadow}
|\cO| &\equiv \frac 1 {\cN_\cO} \int D^d X |\cO(X)\>\<\tl\cO(X)|
\end{align}
is a projector onto the conformal multiplet of $\cO$.  The object $|\cO|$, inserted within a correlator $\<\f_1\dots \f_m \f_{m+1}\dots\f_n\>$, is shorthand for the conformal integral of a product of correlators, supplemented by appropriate monodromy projections
\begin{align}
\<\f_1\dots\f_m|\cO|\f_{m+1}\dots\f_n\> &\equiv \frac 1 {\cN_\cO}\left.\int D^d X \<\f_1\dots\f_m\cO(X)\>\<\tl\cO(X)\f_{m+1}\dots\f_n\>\right|_{M=e^{2\pi i \vf}}.
\end{align}
Here $M$ maps $X_{ij}\to e^{4\pi i} X_{ij}$ for $i,j\leq m$, and leaves the other $X_{ij}$ invariant.  Consistency with the OPE requires $\vf=\De-\sum_{i\leq m} \De_i$.  (For example, we should project $W_\cO(X_i)$ in~(\ref{eq:opshadowopintegral}) onto the subspace with $M=e^{2\pi i(\De-\De_1-\De_2)}$.  Because of the prefactors in~(\ref{eq:conformalblockdefinition}), this means projecting $F(X_i)$ onto $M=e^{2\pi i \De}$ as in eq.~(\ref{eq:conformalblockfinalexpression}).)
The notion of $|\cO|$ as a projection operator is somewhat formal, since the precise form of the monodromy projection depends on what correlator we are computing.

The constant $\cN_\cO$ can be fixed by demanding that $|\cO|$ act trivially when inserted within a correlator involving $\cO$,
\begin{align}
\<\cO(X)|\cO|\dots\> &= \frac 1 {\cN_\cO}\int D^d X_1 \<\cO(X)\cO(X_1)\>\frac{D^d X_0}{(-2 X_1\.X_0)^{d-\De}}\<\cO(X_0)\dots\>
\\
&= \frac 1 {\cN_\cO}\int \frac{D^d X_1}{(-2X\.X_1)^\De}\frac{D^d X_0}{(-2 X_1\.X_0)^{d-\De}}\<\cO(X_0)\dots\>\\
&\stackrel{?}=\<\cO(X)\dots\>.
\label{eq:conditionfornormalization}
\end{align}
Fortunately, we can determine $\cN_\cO$ from this condition without too much computation.  Note that any correlator $\<\cO(X_0)\dots\>$ can be written as a linear combination of functions
\begin{align}
\label{eq:basisforscalarcorrelators}
\frac{1}{(-2X_0\.Y)^\De},
\end{align}
where $Y$ is some (not necessarily null) vector.  For instance, we may combine denominators using Feynman parameters, so that $Y$ is a combination of parameters and other points in the correlator.\footnote{Feynman parameterization is singular for numerator factors with positive integer powers $(-2X\.X_i)^n$.  For the argument here, one should regulate these singularities by taking $n\to n+\e$.}  The power of $X_0$ is fixed by homogeneity.  We have
\begin{align}
\int \frac{D^d X_0}{(-2X_1\.X_0)^{d-\De}}\frac{1}{(-2X_0\.Y)^\De} 
&=\frac{\pi^h \G(\De-h)}{\G(\De)}\frac{(-Y^2)^{h-\De}}{(-2X_1\.Y)^{d-\De}},
\label{eq:intermediateresult}
\end{align}
where $h\equiv d/2$.\footnote{For $|\cO|$ inserted within a two-point function $\<\cO(X)|\cO|\cO(Y)\>$, the vector $Y$ is null and this intermediate result is singular.  Taking $Y$ slightly off the null-cone provides a regularization.}
Iterating this formula a second time with $\De\to d-\De$ gives
\begin{align}
\int  \frac{ D^d X_1}{(-2X_2\.X_1)^\De}\frac{D^d X_0}{(-2X_1\.X_0)^{d-\De}}\frac{1}{(-2X_0\.Y)^\De}
&=\frac{\pi^d \G(\De-h)\G(h-\De)}{\G(\De)\G(d-\De)}\frac{1}{(-2X_2\.Y)^\De}.
\end{align}
This result has exactly the same form as our starting point~(\ref{eq:basisforscalarcorrelators}), up to a $Y$-independent constant.  Taking linear combinations for different $Y$, it follows that
\begin{align}
\<\cO(X)|\cO|\dots\> &= \frac 1 {\cN_\cO} \frac{\pi^d \G(\De-h)\G(h-\De)}{\G(\De)\G(d-\De)} \<\cO(X)\dots\>,
\end{align}
so we should choose 
\begin{align}
\cN_\cO &=  \frac{\pi^d \G(\De-h)\G(h-\De)}{\G(\De)\G(d-\De)}.
\end{align}

Our strategy for computing higher-spin conformal blocks will be to find conformally-invariant projectors analogous to (\ref{eq:projectionoperatorintermsofshadow}) for operators in nontrivial Lorentz representations.  Inserting the projector within a four-point function, we obtain expressions for conformal partial waves in terms of monodromy-projected conformal integrals.  We give further details in section~\ref{sec:higherspinblocksgeneralmethod}.  For now, let us turn to actually computing those integrals.  

\section{Conformal Integrals and Monodromy Invariants}
\label{sec:conformalintegrals}

\subsection{Scalar Four-point Integrals}

As we saw in the previous section, the conformal block for scalar exchange in a four-point function $\<\f\f\f\f\>$ depends on the monodromy-projected conformal integral
\begin{align}
F(X_i)|_{M=e^{2\pi i \De}} &\propto \left.\int D^d X_0 \frac{1}{X_{10}^{\De/2}X_{20}^{\De/2}X_{03}^{(d-\De)/2}X_{04}^{(d-\De)/2}}\right|_{M=1} \x X_{12}^{\De/2}X_{34}^{(d-\De)/2},
\label{eq:integralwedliketocompute}
\end{align}
where $M:X_{12}\to e^{4\pi i} X_{12}$.  Note that since $X_{12}^{\De/2}$ already has the correct eigenvalue $e^{2\pi i (\vf+\De_1+\De_2)}=e^{2\pi i\De}$ under $M$, we would like the $M=1$ subspace of the four-point integral above.  In this section, we will give a simple computation of the above quantity and its generalizations.\footnote{Restricting to the Poincar\'e section $X^+=1$, conformal 2-, 3-, and $4$-point integrals become generalized bubble, triangle, and box integrals, and our results are consistent with known results in those cases.  For recent computations of these integrals in Mellin space, see \cite{Paulos:2012nu}.}
  In even dimensions, the result can be cast in terms of elementary hypergeometric functions, reproducing expressions in \cite{DO1,DO2}.  Unlike the derivation in \cite{DO1,DO2}, ours easily generalizes to the case of conformal integrals with tensor indices, which will be needed to compute conformal blocks for operators with spin.

In more general computations, explicit factors of $X_{12}$ will again have the correct eigenvalue under monodromy, as for $X_{12}^{\De/2}$ in~(\ref{eq:integralwedliketocompute}).  Stripping them off, we will be left with the problem of computing the $M=1$ projection of the four-point integral
\begin{align}
\label{eq:basicfourpointintegral}
I(X_i) &= \frac{\G(a)\G(b)\G(e)\G(f)}{\pi^h \G(h)} \int \frac{D^{2h} X_0}{X_{10}^{a}X_{20}^{b}X_{03}^{e}X_{04}^{f}},
\end{align}
where $d=2h$ is the dimension of spacetime, and $a+b+e+f=2h$ so that the projective measure is well-defined.  The constants out front are chosen for later convenience.  We will assume $X_{ij}>0$.

To begin, combine denominators with the Feynman/Schwinger parameterization~(\ref{eq:schwingerformula}) and apply~(\ref{eq:basicconformalintegralanswer}) to obtain
\begin{align}
I(X_i) 
 =& \int_0^\oo \frac{d\b}{\b}\frac{d\g}{\g}\frac{d\de}{\de} \frac{\b^{b}\g^{e}\de^{f}}{(-(X_1+\b X_2+\g X_3+\de X_4)^2)^{h}}\equiv I_{b,e,f}^{(h)}(X_i)
 \label{eq:basicscalarintegral}
 \\
 =& \frac{\G(h-f)\G(f)}{\G(h)}
\int_0^\oo \frac{d\b}{\b}\frac{d\g}{\g} \frac{\b^{b}\g^{e}}{(\b X_{12}+\g X_{13}+\b\g X_{23})^{h-f}(X_{14}+\b X_{24}+\g X_{34})^{f}},
\label{eq:monodromyready}
\end{align}
where in the last line we have performed the integral over $\de$.  We denote the integral~(\ref{eq:basicscalarintegral}) as $I^{(h)}_{b,e,f}(X_i)$ for convenience in later sections.  

\begin{figure}[h!]
\begin{center}
\begin{psfrags}
\psfrag{0}[B][B][1][0]{$0$}
\psfrag{1}[B][B][1][0]{$\g_1$}
\psfrag{2}[B][B][1][0]{$\g_2$}
\psfrag{i}[B][B][1][0]{$\oo$}
\psfrag{J}[B][B][1][0]{$I_1$}
\psfrag{K}[B][B][1][0]{$I_2$}
\psfrag{L}[B][B][1][0]{$I_3$}
\psfrag{g}[B][B][1][0]{$\g$}
\includegraphics[width=100mm]{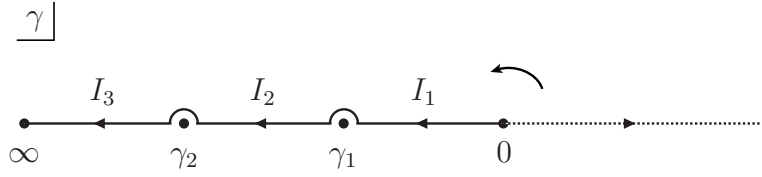}
\end{psfrags}
\end{center}
\caption{We rotate the contour for the integral~(\ref{eq:monodromyready}) in the complex $\g$-plane so that it passes along the negative real axis.  It breaks up into $I_1,I_2,I_3$ as shown.}
\label{fig:initialcontourchoice}
\end{figure}

Let us clarify the analytic structure of~(\ref{eq:monodromyready}).  With $\b$ fixed, the integral over $\g$ traces a path on a multi-sheeted cover $\Sigma\to\P^1$, with branch points at
\begin{align}
0,\qquad \g_1 &\equiv -\frac{\b X_{12}}{X_{13}+\b X_{23}},\qquad
\g_2 \equiv -\frac{X_{14}+\b X_{24}}{X_{34}},\qquad \textrm{and } \oo.
\end{align}
Note that for sufficiently small $X_{12}$, we have $|\g_1| < |\g_2|$, independent of the value of $\b$.  We may deform the $\g$ contour as depicted in figure~\ref{fig:initialcontourchoice}, so that it follows the negative real axis, moving above $\g_1$ and $\g_2$.  Our $\g$-integral~(\ref{eq:monodromyready}) can thus be written
\begin{align}
I &= I_1 + I_2 + I_3,
\end{align}
where $I_1, I_2, I_3$ are integrals along the intervals $[0,\g_1], [\g_1,\g_2]$, and $[\g_2,\oo]$, respectively.\footnote{These integrals may have power-law singularities $\g^{-x}$ with non-integral $x$ at their endpoints.  We define them by analytic continuation in $x$.}

\begin{figure}[h!]
\begin{center}
\begin{psfrags}
\psfrag{J}[B][B][1][0]{$I_1$}
\psfrag{K}[B][B][1][0]{$I_2$}
\psfrag{L}[B][B][1][0]{$I_3$}
\psfrag{M}[B][B][1][0]{$-e^{i\f_\oo}I_3 $}
\psfrag{N}[B][B][1][0]{$-e^{i\f_\oo+i\f_2}I_2$}
\psfrag{P}[B][B][1][0]{$e^{-i\f_0}I_1$}
\psfrag{g}[B][B][1][0]{$\g$}
\psfrag{Z}[B][B][1][0]{$=\ \ \ \ 0$}
\includegraphics[width=85mm]{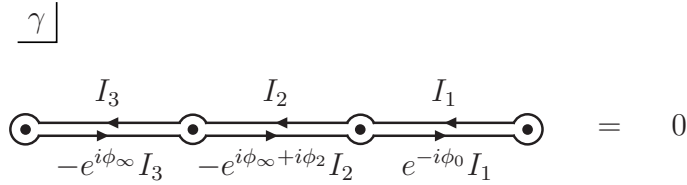}
\end{psfrags}
\end{center}
\caption{This contour is contractible on the punctured Riemann sphere, and thus integrates to zero.  However, it can also be written as a linear combination of the segments $I_1,I_2,I_3$ as shown.  The associated phases are determined by the monodromy around the branch points $0,\g_1,\g_2,\oo$.}
\label{fig:contractiblecontour}
\end{figure}

The integrals $I_1,I_2,I_3$ are not linearly independent.  A contour encircling all four branch points $0,\g_1,\g_2,\oo$ is contractible on $\Sigma$, so integrates to zero.  On the other hand, such a contour can be deformed to a linear combination of $I_1,I_2,I_3$ as shown in figure~\ref{fig:contractiblecontour}, so that
\begin{align}
\label{eq:lineardependenceamongcontours}
0 &= I_1(1-e^{-i\f_0})+I_2(1-e^{i\f_\oo+i\f_2})+I_3(1-e^{i\f_\oo}),
\end{align}
where
\begin{align}
\f_0\equiv 2\pi e,\quad\f_1\equiv 2\pi(f-h) ,\quad \f_2\equiv -2\pi f,\quad\f_\oo\equiv 2\pi (h-e)
\end{align}
are the phases associated with moving counterclockwise around each branch point.  From~(\ref{eq:lineardependenceamongcontours}) we can solve for $I_2$ and express $I$ in terms of $I_1$ and $I_3$,
\begin{align}
\label{eq:eigenspacedecomposition}
I &= \frac{e^{-i\f_0}-e^{i\f_\oo+i\f_2}}{1-e^{i\f_\oo+i\f_2}}I_1 + \frac{e^{i\f_\oo}-e^{i\f_\oo+i\f_2}}{1-e^{i\f_\oo+i\f_2}}I_3.
\end{align}

\begin{figure}[h!]
\begin{center}
\begin{psfrags}
\psfrag{J}[B][B][1][0]{$I_1$}
\psfrag{K}[B][B][1][0]{$I_2$}
\psfrag{L}[B][B][1][0]{$I_3$}
\psfrag{0}[B][B][1][0]{$0$}
\psfrag{1}[B][B][1][0]{$\g_1$}
\psfrag{2}[B][B][1][0]{$\g_2$}
\psfrag{i}[B][B][1][0]{$\oo$}
\psfrag{g}[B][B][1][0]{$\g$}
\psfrag{M}[B][B][1][0]{$M$}
\includegraphics[width=90mm]{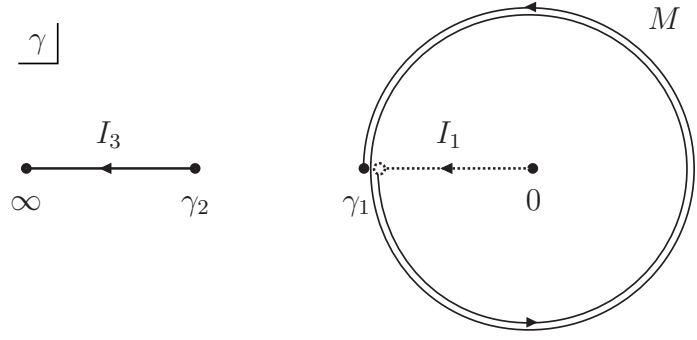}
\end{psfrags}
\end{center}
\caption{The monodromy $M$ moves $\g_1$ twice around the origin, so that $I_1$ picks up a phase $e^{2i\f_0}$, while $I_3$ remains invariant.}
\label{fig:monodromyaction}
\end{figure}

We are finally ready to understand the behavior of $I(X_i)$ under monodromy $M:X_{12}\to e^{4\pi i}X_{12}$.  $M$ moves the branch point $\g_1$ twice around the origin (figure~\ref{fig:monodromyaction}), so that the integral $I_1$ picks up a phase $e^{2i\f_0}$.  On the other hand, $M$ leaves the integral $I_3$ invariant, since neither $\g_2$ nor $\oo$ moves, and $\g_1$ does not pass through the $I_3$ integration contour.  In other words,~(\ref{eq:eigenspacedecomposition}) is precisely the decomposition of $I$ into eigenvectors of $M$.  The monodromy-invariant component is
\begin{align}
I|_{M=1} &= \frac{e^{i\f_\oo}-e^{i\f_\oo+i\f_2}}{1-e^{i\f_\oo+i\f_2}}I_3
=e^{i\pi (h-e)}\frac{\sin(\pi f))}{\sin(\pi (e+f-h))} I_3.
\end{align}

Having identified the correct monodromy-invariant contour, let us change variables in~(\ref{eq:monodromyready}) to $\b\to\frac{X_{14}}{X_{24}}\b$, $\g\to e^{i\pi}\frac{X_{14}}{X_{34}}\g$.  This maps $\g_2\to \b+1$ and gives
\begin{align}
\label{eq:intermediateintegral1}
I|_{M=1} =&
\frac{\G(h-f)\G(f)}{\G(h)}\frac{\sin(\pi f)}{\sin(\pi (e+f-h))}X_{14}^{b+e-h}X_{13}^{f-h}X_{34}^{h-f-e}X_{24}^{-b}\nn\\
&\x\int_0^\oo \frac{d\b}{\b}\int_{\b+1}^\oo\frac{d\g}{\g} \frac{\b^{b}\g^{e}}{(\g+v\b\g-u\b)^{h-f}(\g-\b-1)^{f}}.
\end{align}
It's now straightforward to expand the denominator and evaluate the integral as a power series in $u$ and $1-v$.

When the dimension of spacetime is even, so that $h$ is an integer, we can proceed further.  The computation is easiest when the exponents in the denominator sum to $1$, so let us bring~(\ref{eq:intermediateintegral1}) to this form:
\begin{align}
I|_{M=1} &=
\frac{\pi}{\G(h)\sin(\pi(e+f-h))}X_{14}^{b+e-h}X_{13}^{f-h}X_{34}^{h-f-e}X_{24}^{-b}\nn\\
&\x\p{-\pdr{}{v}}^{h-1}\int_0^\oo \frac{d\b}{\b}\int_{\b+1}^\oo\frac{d\g}{\g} \frac{\b^{b-h+1}\g^{e-h+1}}{(\g+v\b\g-u\b)^{1-f}(\g-\b-1)^{f}}\qquad(h\in\Z).
\end{align}
Finally, write $u=z\bar z, v=(1-z)(1-\bar z)$, and make the change of variables
\begin{align}
\label{eq:magicvariables}
\b=\frac{s}{(1-s)(1-t\bar z)},\quad \g=\frac{1}{(1-t)(1-s)},\qquad s,t\in[0,1].
\end{align}
Our expression factorizes into a product of one-dimensional integrals which produce ${}_2 F_1$ hypergeometric functions of $z$ and $\bar z$,
\begin{align}
I|_{M=1} =& \frac{\G(2h-b-e-f)\G(1+b-h)\G(1-f)\G(h-e)\G(e+f-h)}{\G(h)\G(1+h-e-f)}X_{14}^{b+e-h}X_{13}^{f-h}X_{34}^{h-f-e}X_{24}^{-b}\nn\\
&\x\p{-\pdr{}{v}}^{h-1}F(z)F(\bar z)\qquad(\textrm{even dimensions}, h\in\Z)\nn\\
F(x)\equiv& {}_2F_1(b+1-h,1-f,1+h-e-f,x).
\label{eq:evendimresult}
\end{align}
In terms of $z$ and $\bar z$, the derivative operator reads
\begin{align}
-\pdr{}{v} &= \frac 1 {z-\bar z}\p{z\pdr{}{z}-\bar z \pdr{}{\bar z}}.
\end{align}
When $h\notin\Z$, the change of variables~(\ref{eq:magicvariables}) does not factorize the integral~(\ref{eq:intermediateintegral1}), but instead gives a result which can be expanded as a series of hypergeometric functions.

With~(\ref{eq:evendimresult}) we can give compact expressions for scalar conformal blocks in  even dimensions.  The monodromy projection of the right hand side of~(\ref{eq:opshadowopintegral}) gives the conformal partial wave for exchange of a scalar $\cO$ with dimension $\De$ between scalars $\f_i$ with dimensions $\De_i$.  The conformal block is given by
\begin{align}
g_\cO(X_i) =& X_{12}^{\frac{\De_1+\De_2}{2}}X_{34}^{\frac{\De_3+\De_4}{2}}\p{\frac{X_{13}}{X_{14}}}^{\frac{\De_{34}}{2}}\p{\frac{X_{14}}{X_{24}}}^{\frac{\De_{12}}{2}}\nn\\
&\x
\left.\int D^d X_0\<\f_1(X_1)\f_2(X_2)\cO(X_0)\>\<\tl\cO(X_0)\f_3(X_3)\f_4(X_4)\>\right|_{M=e^{2\pi i\De}}\\
=&\frac{\p{\frac{\De-\De_{12}}{2}}_{1-h}\p{\frac{\De+\De_{34}}{2}}_{1-h}}{(\De)_{1-h}}\p{-\pdr{}{v}}^{h-1}k(z)k(\bar z) \nn\\
k(x)\equiv& x^{\De/2}{}_2F_1\p{\frac{\De-\De_{12}}{2}+1-h,\frac{\De+\De_{34}}{2}+1-h,\De+1-h,x},
\end{align}
where $\De_{ij}\equiv \De_i-\De_j$, and $(a)_n=\G(a+n)/\G(a)$ is the Pochhammer symbol.  This agrees with the results of \cite{DO1,DO2} for $2,4$, and $6$ dimensions, after applying elementary hypergeometric function identities.  In particular, eq.~(\ref{eq:DOscalarblock}) for $d=4$, $\ell=0$ is easily verified.

\subsection{Tensor Four-point Integrals}

These results generalize straightforwardly to conformal integrals with nontrivial tensor structure.  The most general possible four-point integral is
\begin{align}
\label{eq:basictensorfourptintegrals}
I^{m_1\dots m_n}(X_i) &=
\frac{\G(a)\G(b)\G(e)\G(f)}{\pi^h \G(h+n)}\int D^{2h}X_0 \frac{X_0^{m_1}\cdots X_0^{m_n}}{X_{10}^{a}X_{20}^{b}X_{03}^{e}X_{04}^{f}},
\end{align}
where now $a+b+e+f=2h+n$.  Combining denominators and and applying~(\ref{eq:conformalintegralwithindices}), we obtain
\begin{align}
I^{m_1\dots m_n}(X_i) =& \int_0^\oo \frac{d\b}{\b}\frac{d\g}{\g}\frac{d\de}{\de} \b^b\g^e\de^f  \frac{X_{\b,\g,\de}^{m_1}\cdots X_{\b,\g,\de}^{m_n}}{(-X_{\b,\g,\de}^2)^{n+h}}-\textrm{traces}\\
X_{\b,\g,\de}\equiv& X_1+\b X_2+\g X_3+\de X_4.
\end{align}
Finally, expanding the numerator in monomials, we can evaluate the result in terms of the scalar integrals~(\ref{eq:basicscalarintegral}),
\begin{align}
\label{eq:basictensorfourptresult}
I^{m_1\dots m_n}(X_i) =&
\sum_{p+q+r+s=n}\frac{n!}{p!q!r!s!}I^{(h+n)}_{b+q,e+r,f+s}(X_i)\nn\\
&\quad\x X_1^{(m_1}\cdots X_1^{m_p}X_2^{m_{p+1}}\cdots X_2^{m_{p+q}} X_3^{m_{p+q+1}}\cdots X_3^{m_{n-s}}X_4^{m_{n-s+1}}\cdots X_4^{m_n)}\nn\\
&\quad-\textrm{traces}.
\end{align}
Since the $X_i$ prefactors have trivial monodromy, projection onto the monodromy-invariant subspace can be performed termwise on each scalar integral $I_{b,e,f}^{(h)}$.

\section{Higher Spin Conformal Blocks}
\label{sec:higherspinconformalblocks}

\subsection{General Method}
\label{sec:higherspinblocksgeneralmethod}

With the language of section~\ref{sec:constructingconformalblocks} and the results of section~\ref{sec:conformalintegrals}, computing higher-spin conformal blocks is a simple generalization of the case for scalar blocks.  Consider a four-point function of primary operators $\phi_i$ in different Lorentz representations.  The first step is to lift the operators $\f_i$ to embedding space fields $\phi_i^{I_i}(X_i)$, where $I_i$ is a general embedding space Lorentz index.  The precise way to lift $\f_i$ depends on its Lorentz representation and the spacetime dimension.  We will give several concrete examples below.

Three-point functions of $\f_i$'s with an operator $\cO^J(X)$ and its conjugate $\bar \cO_J(X)$ are generically a sum of several tensor structures, each with its own independent OPE coefficient,
\begin{align}
\label{eq:generalthreeptfunction1}
\<\cO^J\f_1^{I_1}\f_2^{I_2}\> &= \<\cO^J\f_1^{I_1}\f_2^{I_2}\>^{(m)}\l_m,\\
\label{eq:generalthreeptfunction2}
\<\bar\cO_J\f_3^{I_3}\f_4^{I_4}\> &= \<\bar\cO_J\f_3^{I_3}\f_4^{I_4}\>^{(n)}\eta_n.
\end{align}
Here, we have denoted the independent structures by a superscript $\<\cdots\>^{(m)}$, the associated OPE coefficients by $\l_m$ and $\eta_n$, and a sum over $m,n$ is implied.  The number of structures in each three-point function depends on the Lorentz representations of $\cO$ and $\f_i$.

The four-point function $\<\f_1\f_2\f_3\f_3\>$ has a conformal partial wave expansion
\begin{align}
\<\f_1^{I_1}(X_1)\f_2^{I_2}(X_2)\f_3^{I_3}(X_3)\f_4^{I_4}(X_4)\>
&=
\sum_{\cO\in\f_1\x\f_2} \l_m \eta_n W^{(m,n)I_1I_2I_3I_4}_\cO(X_i),
\end{align}
where $W^{(m,n)}_\cO$ is the conformal partial wave corresponding to the pair of tensor structures $(m,n)$.  To compute $W^{(m,n)}_\cO$, we need a conformally invariant projector $|\cO|$ analogous to~(\ref{eq:projectionoperatorintermsofshadow}) which enables us to ``sew together" the three-point functions~(\ref{eq:generalthreeptfunction1}) and (\ref{eq:generalthreeptfunction2}).

In the cases we will encounter below, the embedding space lift of $\cO^J$ will have gauge-redundancies, which $|\cO|$ must respect.  Our projector will have the general form
\begin{align}
\label{eq:generalprojectorform}
|\cO| &=
\int D^d X\,D^d Y|\cO^J(X)\>\frac{\Pi(X,Y)_J^K}{(-2X\.Y)^{d+\deg \cO+\deg \Pi}}\<\bar\cO_K(Y)|,
\end{align}
where $\Pi(X,Y)$ is a tensor built from $X,Y$ that ensures gauge-invariance, and the denominator is chosen so that the projective integral is well-defined.  Specifically, $\deg\cO$ is the degree of the embedding-space lift of $\cO$, and $\deg\Pi$ is the degree of $\Pi(X,Y)$ in either $X$ or $Y$ (which must be the same).  After performing the integral, we must additionally project out the shadow contribution.  The integrals we encounter in practice will always be reducible to a sum of basic tensor four-point integrals (\ref{eq:basictensorfourptintegrals}) whose monodromy projections we can evaluate with (\ref{eq:basictensorfourptresult}) and (\ref{eq:intermediateintegral1}).

Inserting $|\cO|$ within a four-point function is guaranteed by conformal invariance to produce a linear combination of conformal partial waves for the exchange of $\cO$.  To normalize them correctly, we should insert $|\cO|$ within a three-point function, as in~(\ref{eq:conditionfornormalization}).  In general, the projector can mix different tensor structures,
\begin{align}
\<\bar\cO_J|\cO|\f_3^{I_3}\f_4^{I_4}\> &= \<\bar\cO_J\f_3^{I_3} \f_4^{I_4}\>^{(m)}(\cM_{\cO34})_m{}^n\eta_n.
\end{align}
Thus, to obtain the conformal partial waves corresponding to a specific pair of tensor structures, we should multiply by the inverse of the mixing matrix $\cM_{\cO34}$,
\begin{align}
\label{eq:generalconformalpartialwave}
W^{m,n}_\cO &= {}^{(m)}\<\f_1\f_2|\cO|\f_3\f_4\>^{(k)}(\cM_{\cO34}^{-1})_k{}^n.
\end{align}
With this prescription, $W_\cO^{m,n}$ has the correct limiting behavior as $x_1\to x_2$ (and $\f_1\f_2$ becomes better approximated by linear combinations of $\bar\cO$).  Since conformal partial waves are determined by either of the limits $x_1\to x_2$ or $x_3\to x_4$, the apparent asymmetry of (\ref{eq:generalconformalpartialwave}) under $1,2\leftrightarrow 3,4$ is illusory.  Indeed, we must also have
\begin{align}
\label{eq:generalconformalpartialwave2}
W^{m,n}_\cO &= (\cM_{\bar\cO 12}^{-1})_k{}^m{}^{(k)}\<\f_1\f_2|\cO|\f_3\f_4\>^{(n)}.
\end{align}
In the examples below, the mixing matrix will be an overall constant, so the equivalence between (\ref{eq:generalconformalpartialwave}) and (\ref{eq:generalconformalpartialwave2}) will be obvious.\footnote{It may be possible to show that this is always true, analogous to the arguments for scalar $\cO$ given in section~\ref{sec:projectorsandshadows}.  It would follow if $|\cO|$ can be interpreted as a projection operator on a fixed Hilbert space in radial quantization.  The fact that the monodromy $M$ depends on the positions $X_{1},X_2$ makes such an interpretation difficult.}

\subsection{Tensor Operators in the Embedding Space}
\label{sec:spinningcorrelatorsformalism}

The simplest operators to which we can apply this machinery are tensors.  In this section, we focus on traceless tensors
$\f^{\mu_1\dots\mu_\ell}$ whose Lorentz representations are specified by some pattern of symmetries in their indices.
This is sufficient for understanding all bosonic operators in 3D CFTs, since these can always be decomposed into traceless symmetric representations of the Lorentz group.  In higher than three dimensions, such tensors could be reducible (for instance, an antisymmetric tensor in four dimensions can be decomposed into anti-/self-dual parts), and it is convenient to use more refined techniques.  We will develop them for 4d CFTs in section~\ref{sec:twistormethods}.

As argued in the previous section, two elements are required to compute conformal blocks for tensor operators: 1) a way to lift tensors to the embedding space, and 2) a gauge- and conformally-invariant projector.  Embedding space lifts for tensors were introduced in \cite{Ferrara:1973eg} and further developed in \cite{Weinberg:2010fx,Costa:2011mg}. A primary operator $\f^{\mu_1\dots\mu_\ell}(x)$ with dimension $\De$ transforming as a traceless tensor of the Lorentz group can be lifted to an embedding space tensor $\Phi^{m_1\dots m_\ell}(X)$ with the following properties:
\begin{enumerate}
\item defined on the null-cone,
\item traceless and possessing the same index symmetries as $\f^{\mu_1\dots\mu_\ell}$, \label{prop:traceless}
\item defined modulo tensors of the form $X^{m_i}\Lambda^{m_1\dots\widehat m_i\dots m_\ell}(X)$, \label{prop:gaugeinvariance}
\item transverse $X_{m_i}\Phi^{m_1\dots m_i\dots m_\ell}(X)=0$, \label{prop:transverseness}
\item degree $-\De$ in $X$.
\end{enumerate}
One can recover the original tensor $\f^{\mu_1\dots\mu_\ell}(x)$ by restricting to the Poincar\'e section $X^m=(1,x^2,x^\mu)$, and projecting indices as follows
\begin{align}
\f_{\mu_1\dots\mu_\ell}(x) &= \frac{\ptl X^{m_1}}{\ptl x^{\mu_1}}\cdots \frac{\ptl X^{m_\ell}}{\ptl x^{\mu_\ell}}\Phi_{m_1\dots m_\ell}(X).
\end{align}

When $\Phi^{m_1\dots m_\ell}$ is symmetric in its indices (so that it transforms in a spin-$\ell$ representation of the Lorentz group), it is often convenient to use index-free notation.  We introduce an auxiliary vector $Z_m$, and form the contraction
\begin{align}
\label{eq:indexfreetensornotation}
\Phi(X,Z) &\equiv \Phi^{m_1\dots m_\ell}(X)Z_{m_1}\dots Z_{m_\ell},
\end{align}
which is a homogeneous polynomial of degree $\ell$ in $Z$.  The components of $\Phi^{m_1\dots m_\ell}$ can be recovered by taking derivatives with respect to $Z$,
\begin{align}
\label{eq:gettingbacktoindices}
\Phi^{m_1\dots m_\ell}(X) &=\frac 1 {\ell!}\pdr{}{Z_{m_1}}\cdots \pdr{}{Z_{m_\ell}} \Phi(X,Z)-\mathrm{traces}.
\end{align}
Each property of $\Phi^{m_1\dots m_\ell}$ is reflected in properties of $\Phi(X,Z)$.  The tracelessness condition (\ref{prop:traceless}) means that we can restrict $\Phi(X,Z)$ to the null-cone $Z^2=0$ without losing any information.  To apply~(\ref{eq:gettingbacktoindices}), we must then extend $\Phi(X,Z)$ away from $Z^2=0$.  Any two extensions will differ by an amount proportional to $Z^2$, which vanishes in~(\ref{eq:gettingbacktoindices}) after subtracting traces.   The redundancy (\ref{prop:gaugeinvariance}) means that we can further restrict $\Phi(X,Z)$ to the plane $Z\.X=0$.  To apply~(\ref{eq:gettingbacktoindices}), we must extend $\Phi(X,Z)$ away from this plane, and different extensions lead to gauge-equivalent tensors.  Finally,
transverseness (\ref{prop:transverseness}) implies that $\Phi(X,Z)$ has a gauge-redundancy under $Z\to Z+\l X$, for $\l\in\R$.

The advantage of index-free notation is that complicated conformally-covariant tensors can become simple algebraic expressions in terms of conformal invariants.  Correlators of symmetric tensors $\Phi_i(X_i,Z_i)$ must be gauge- and conformally-invariant functions of $X_i$ and $Z_i$ with the correct homogeneity properties.  In two- and three-point correlators, such functions can be constructed as polynomials in the basic invariants
\begin{align}
\label{eq:tensortwoandthreeptstructures}
V_{i,jk} \equiv \frac{X_j\.Z_i}{X_{ij}}-\frac{X_k\.Z_i}{X_{ik}},
\qquad
H_{ij} \equiv \frac{(X_i\.X_j)(Z_i\.Z_j)-(X_i\.Z_j) (X_j\.Z_i)}{X_i\.X_j},
\end{align}
along with the $X_{ij}$.  In 3d and 4d, other invariants involving $\e$-tensors are possible \cite{Costa:2011mg}.

\subsection{Projectors for Tensor Operators}

Given a tensor operator $\cO^{m_1\dots m_\ell}$, there is an essentially unique projector $|\cO|$ of the form (\ref{eq:generalprojectorform}) compatible with all of the above properties,
\begin{align}
\label{eq:spinprojector1}
|\cO| =& \int D^d X D^d Y |\cO_{m_1\dots m_\ell}(X)\>\frac{\prod_i(\eta^{m_i n_i}(X\.Y)-Y^{m_i}X^{n_i})}{(-2 X\.Y)^{d-\De+\ell}}\<\cO_{n_1\dots n_\ell}(Y)|.
\end{align}
The tensors $\eta^{m_i n_i}(X\.Y)-Y^{m_i}X^{n_i}$ are required to ensure invariance under gauge transformations $\cO_{n_i\dots}(X)\to \cO_{n_i\dots}(X)+X_{n_i}\L_{\dots}$.

To render~(\ref{eq:spinprojector1}) in a form more similar to~(\ref{eq:projectionoperatorintermsofshadow}), we can define the shadow operator
\begin{align}
\label{eq:tensorshadowtransform}
\tl \cO(X)^{m_1\dots m_\ell} &\equiv \int D^d Y \frac{\prod_i(\eta^{m_i n_i}(X\.Y)-Y^{m_i}X^{n_i})}{(-2 X\.Y)^{d-\De+\ell}}\cO_{n_1\dots n_\ell}(Y).
\end{align}
In terms of $\tl\cO$, the projector $|\cO|$ becomes simply
\begin{align}
|\cO| &= \int D^d X |\cO_{m_1\dots m_\ell}(X)\>\<\tl \cO^{m_1\dots m_\ell}(X)|.
\end{align}

Using index-free notation for symmetric tensors, the shadow operator can be written
\begin{align}
\label{eq:spinshadow}
\tl \cO(X,Z) &= \int D^d Y \frac{1}{(-2 X\.Y)^{d-\De+\ell}}\cO(Y,C_{ZX}\.Y),
\end{align}
where $C_{ZX}^{mn}\equiv Z^m X^n-X^m Z^n$.  Note that $\tl \cO$ is well-defined, since $Y\. C_{ZX}\.Y=0$ and $(C_{ZX}\.Y)^2=0$ (assuming that $Z\.X=0$).  Further, $\tl \cO(X,Z)$ automatically enjoys the correct gauge redundancy, since $C_{ZX}$ is invariant under $Z\to Z+\l X$s.  Finally, since $\tl\cO(X,Z)$ has degrees $-(d-\De)$ and $\ell$ in $X$ and $Z$, it formally possesses all the required properties of a primary operator with dimension $d-\De$ and spin $\ell$.  

Before moving on to examples, let us quickly summarize the approach of \cite{Costa:2011mg} for computing conformal blocks of symmetric tensors.  The authors define differential operators $\cD^{(m)}_\mathrm{left}$ and $\cD^{(n)}_\mathrm{right}$ that turn three-point functions of scalars $\vf_i$ into three-point functions of higher-spin operators $\f_i$,
\begin{align}
{}^{(m)}\<\f_1(X_1,Z_1)\f_2(X_2,Z_2)\cO(X,Z)\> &= \cD^{(m)}_\mathrm{left}\<\vf_1(X_1)\vf_2(X_2)\cO(X,Z)\>\\
{}^{(n)}\<\f_3(X_3,Z_3)\f_4(X_4,Z_4)\cO(X,Z)\> &= \cD^{(n)}_\mathrm{right}\<\vf_3(X_3)\vf_4(X_4)\cO(X,Z)\>.
\end{align}
Here $m,n$ index the possible tensor structures.  $\cD^{(m)}_\mathrm{left}$ and $\cD^{(n)}_\mathrm{right}$ are constructed to involve only the external coordinates $X_i,Z_i$.  By linearity, and the fact that the $\cD$'s act trivially under monodromy, it's clear that
\begin{align}
&{}^{(m)}\<\f_1(X_1,Z_1)\f_2(X_2,Z_2)|\cO|\f_3(X_3,Z_3)\f_4(X_4,Z_4)\>^{(n)}\nn\\
&= \cD^{(m)}_\mathrm{left}\cD^{(n)}_\mathrm{right} \<\vf_1(X_1)\vf_2(X_2)|\cO|\vf_3(X_3)\vf_4(X_4)\>,
\end{align}
so conformal partial waves for $\cO$ exchanged between $\f_i(X_i,Z_i)$ are derivatives of conformal partial waves for $\cO$ exchanged between scalars $\vf_i(X_i)$.  A virtue of this approach is that expressions for lower-spin blocks can be reused in computations of higher-spin blocks.  However, external derivatives cannot change the conformal multiplet of the operator $\cO$ being exchanged.  One must always begin with a ``seed" calculation of some nonzero conformal block involving a given $\cO$.

\subsection{Example: Antisymmetric Tensor Exchange}

The simplest tensor conformal block that is not related via derivatives to a scalar block is the exchange of an antisymmetric tensor $F^{mn}$ in a four-point function of two scalars and two vectors $\<\f_1 J_2^l \f_3 J_4^k\>$.  (In a four-point function with fewer than two vectors, any pairing of the operators would include a pair of scalars.  The OPE of two scalars contains only symmetric tensors, so only these would contribute in the conformal block expansion.)  We work in $d$ dimensions and assume that $F^{mn}$ transforms irreducibly under the Lorentz group.  (Although this is incorrect when $d=4$, the result still applies if the self-dual and anti-self-dual parts of $F$ have the same OPE coefficient in $\f_1\x J_2$ and $\f_3\x J_4$.)

The three-point function $\<F^{mn}\f_3 J_4^l\>$ has a unique allowed tensor structure
\begin{align}
\label{eq:antisymtensorthreeptstructure}
\<F^{mn}(X_0)\f_3(X_3)J_4(X_4,Z_4)\>
&= \frac{((X_0\.X_4)Z_4^m-(X_0\.Z_4) X_4^m)(X_{03}X_4^n-X_{04}X_3^n)-(m\leftrightarrow n)}{
X_{03}^{\frac{\De+\De_3-\De_4+1}{2}}
X_{04}^{\frac{\De+\De_4-\De_3+3}{2}}
X_{34}^{\frac{\De_3+\De_4-\De+1}{2}}
},
\end{align}
where $\De$ is the dimension of $F$, and we are using index-free notation~(\ref{eq:indexfreetensornotation}) for $J_4$.  This expression is fixed by homogeneity and the requirement of transverseness in its indices, up to gauge redundant terms proportional to $X_0^m,X_0^n$, which we have dropped.

From the definition~(\ref{eq:tensorshadowtransform}), we can compute the shadow transform
\begin{align}
\label{eq:antisymtensorshadow}
\<\tl F^{mn}(X_0)\f_3(X_3)J_4(X_4,Z_4)\>
&=\cS_\De\<F^{mn}(X_0)\f_3(X_3)J_4(X_4,Z_4)\>|_{\De\to\tl\De},\\
\textrm{where}\quad\cS_\De
&\equiv
\frac{\pi^h(\De-2)\G(\De-h)}{4\G(\tl\De+1)}
\frac{
\G\big(
\frac{\tl\De+\De_{34}+1}{2}
\big)
\G\big(
\frac{\tl\De-\De_{34}+1}{2}
\big)
}{
\G\big(
\frac{\De+\De_{34}+1}{2}
\big)
\G\big(
\frac{\De-\De_{34}+1}{2}
\big)
}
\end{align}
and $h=d/2$, $\tl\De=d-\De$ as usual.  As expected, the result has the correct form for a three-point function of $\f_3$ and $J_4$ with an antisymmetric tensor of dimension $\tl\De$.

The next step is to determine the mixing matrix $(\cM_{F34})_i{}^j$ from inserting the projector $|F|$ within a three-point function.  Since there is only a single allowed three-point structure, this is just a number $\cM$.  We can compute it with a simple trick.
Suppose $F^{mn}$ is normalized to have two-point function
\begin{align}
\label{eq:antisymtwoptfunction}
\<F^{mn}(X_1)F^{kl}(X_2)\> &=\frac 1 2
\frac{((X_1\.X_2)\eta^{mk}-X_2^mX_1^k)((X_1\.X_2)\eta^{nl}-X_2^nX_1^l)-(m\leftrightarrow n)}{X_{12}^{\De+2}}.
\end{align}
Again, this structure is fixed up to gauge redundancy by homogeneity and transverseness.
Note that the numerator has precisely the same form as the tensor appearing in the definition of $|F|$ (\ref{eq:spinprojector1}).  In fact, inserting $|F|$ between the two-point function~(\ref{eq:antisymtwoptfunction}) and three-point function~(\ref{eq:antisymtensorthreeptstructure}) is equivalent to simply iterating the shadow transform twice. We can read off the result from~(\ref{eq:antisymtensorshadow}),
\begin{align}
\cM\<F\f_3J_4\> &=\<F|F|\f_3J_4\>\nn\\
 &= \<\tl{\tl{F}}\f_3 J_4\>\nn\\
 &= \cS_\De\cS_{\tl\De}\<F\f_3J_4\>\nn\\
 &= \frac{\pi^{2h}(\tl\De-2)(\De-2)\G(\tl\De-h)\G(\De-h)}{16\G(\tl\De+1)\G(\De+1)}\<F\f_3J_4\>.
\end{align}

Applying~(\ref{eq:generalconformalpartialwave}), the conformal partial wave corresponding to exchange of $F^{mn}$ is given by
\begin{align}
&\p{\frac{X_{14}}{X_{13}}}^{\frac{\De_{34}}{2}}\p{\frac{X_{24}}{X_{14}}}^{\frac{\De_{12}}{2}}
\frac{g^{\De_i}_F(u,v)}{X_{12}^{\frac{\De_1+\De_2}{2}}X_{34}^{\frac{\De_3+\De_4}{2}}}
\nn
\\
&=\frac{1}{\cM}\<\f_1(X_1)J_2(X_2,Z_2)|F|\f_3(X_3)J_4(X_4,Z_4)\>\\
&=\frac{1}{\cS_{\tl\De}}\int D^d X \<\f_1(X_1)J_2(X_2,Z_2)F^{mn}(X)\>\p{\left.\<F_{mn}(X)\f_3(X_3)J_4(X_4,Z_4)\>\right|_{\De\to\tl\De}}.
\end{align}

Finally, using~(\ref{eq:basictensorfourptresult}) to evaluate the integral, we obtain\footnote{Many thanks to Miguel Costa and Tobias Hansen for pointing out several typos in a previous version of this expression.}
\begin{align}
g^{\De_i}_F(u,v)&=\frac{2 u^{\De/2-1/2}X_{24}\G(\De+1)}{(2-\tl\De)\G(\a +1)\G(\b +1)\G(\De-\a )\G(\De-\b )\G(h-\De)}
\nn\\
&
\x\Biggl[
V_{2,34} V_{4,23} \left(v \alpha  J_{1,0,1,0}+v (1+\alpha -\Delta ) J_{1,1,1,0}-(v-1) v J_{2,0,1,0}\right)\nn\\
&+V_{2,14} V_{4,23} \left((\alpha
-v \alpha ) J_{1,0,1,0}+(v-1) J_{2,0,0,0}+(v-1) v J_{2,0,1,0}\right)\nn\\
&+V_{2,34} V_{4,12} \left(v (h-1-\beta ) (\Delta-1-\alpha ) J_{0,2,2,1}+v
\alpha  J_{1,0,1,0}+v (1+\alpha -\Delta ) (J_{1,1,1,0}+J_{1,2,1,1})\right.\nn\\
&\quad\quad\quad\quad\quad\left.+v (h+\alpha +\beta -\Delta ) J_{1,1,1,1}+v (1-h+\beta ) J_{1,1,2,1}\right.\nn\\
&\quad\quad\quad\quad\quad\left.-(v-1)v (J_{2,0,1,0}+J_{2,1,1,1})\right)\nn\\
&+V_{2,14} V_{4,12} \left((h+\beta -\Delta ) (\alpha J_{0,1,1,1}-J_{1,1,0,1})+v (h-1-\beta ) ((1+\alpha -\Delta ) J_{0,2,2,1} +J_{1,1,2,1} )\right.\nn\\
&\quad\quad\quad\quad\quad\left.-v\alpha  J_{1,0,1,0}-\alpha  J_{1,0,1,1} +v (\Delta-1-\alpha ) (J_{1,1,1,0}+J_{1,2,1,1})\right.\nn\\
&\quad\quad\quad\quad\quad\left.-2 v (1+\alpha +\beta -\Delta ) J_{1,1,1,1} +(v-1)(v J_{2,0,1,0}+v J_{2,1,1,1}-J_{2,0,0,1})\right)\nn\\
&-\frac{H_{24}}{2 X_{24}} \left(\alpha
 (h+\beta -\Delta ) J_{0,1,1,1}+\alpha  (1-h+\beta ) J_{0,1,2,1}\right.\nn\\
&\quad\quad\quad\left.+(1+\alpha -\Delta ) (h+\beta -\Delta ) J_{0,2,1,1}+v (h-1-\beta ) (\Delta-1-\alpha
) J_{0,2,2,1}\right.\nn\\
&\quad\quad\quad\left.+\frac{1-v}{h+1}(J_{2,0,0,0}+J_{2,0,0,1}+J_{2,0,1,1}+J_{2,1,0,1}+
v J_{2,1,1,1}+uJ_{2,1,1,2})\right)
\Biggr
],
\end{align}
where
\begin{align}
\a & \equiv \frac{\De-\De_{12}-1}{2},\quad
\b \equiv \frac{\De+\De_{34}-1}{2},
\end{align}
$V_{i,jk}, H_{ij}$ are given by~(\ref{eq:tensortwoandthreeptstructures}),
and the $J_{i,j,k,l}$ are shorthand for monodromy-projected conformal four-point integrals,
\begin{align}
J_{i,j,k,l} &\equiv \G(h+i)(X_{14}^{b+e-h-i}X_{13}^{f-h-i}X_{34}^{h+i-f-e}X_{24}^{-b})^{-1} I^{(h+i)}_{b,e,f}|_{M=1},\textrm{ with}\nn\\
b &=\a+i+j-1\nn\\
e &=\b-\Delta+h+i+k-l\nn\\
f &=1-\b+h-k.
\end{align}
The powers of $X_{ij}$ in the definition of $J$ have been chosen so that $J$ is a function of conformal cross-ratios.  In general dimensions, it is given by the expression~(\ref{eq:intermediateintegral1}); when $d$ is even, it can be written in terms of products of hypergeometric functions using~(\ref{eq:evendimresult}).

\section{Twistor Methods for 4d CFTs}
\label{sec:twistormethods}

\subsection{Lifting Spinors to the Embedding Space}

Although the methods of the previous section are sufficient for computations involving tensors, we need a more flexible formalism to deal with more general Lorentz representations.  For the remainder of this work, we focus on CFTs in four dimensions, where twistors provide natural building blocks for conformal invariants.\footnote{In this and subsequent sections, we work with 4d spinors in signature $-+++$.  Conformal integrals can be defined by analytic continuation back to Euclidean signature.  Our conventions for spinors and $\G$-matrices in the embedding space are detailed in Appendix~\ref{app:spinorconventions}.} \footnote{Twistors and supertwistors have been used extensively in the study of superconformal theories, see e.g. \cite{Siegel:1992ic,Siegel:1994cc,Siegel:2010yd,Siegel:2012di}.}

Twistor space $\bT\cong \C^4$ consists of four-component objects
\begin{align}
Z_A &= \left(
\begin{array}{c}
\l_\a\\
\mu^{\dot\a}
\end{array}
\right)
\end{align}
transforming as left-chiral spinors of the conformal group $\SO(4,2)$, or equivalently fundamentals of $\SU(2,2)$.
$\bT$ possesses a totally antisymmetric conformal invariant given by the determinant $\<Z_1Z_2Z_3Z_4\>\equiv\e^{ABCD}Z_{1A}Z_{2B}Z_{3C}Z_{4D}$.  We also have the dual space $\bar\bT$ with coordinates $\bar W^A$, and an invariant pairing $\bar W Z=\bar W^A Z_A$. 

The (complexified) embedding space itself is the antisymmetric tensor-square of twistor space, $\C^6\cong\wedge^2 \bT^4$, and the null-cone consists of precisely the pure tensors (or ``simple bitwistors") under this identification,
\begin{align}
\label{eq:nullconeintermsoftwistors}
X_{AB}&=Z_A W_B - Z_B W_A,
\end{align}
where $X_{AB}\equiv X_m\G^{m}_{AB}$, with $\G^{m}_{AB}$ a chiral gamma-matrix.  In other words, the projective null-cone is isomorphic to the Grassmanian of two-planes in twistor space $\mathrm{Gr}(2,\bT)$.  
Note that the null condition $X^2=0$ implies that $X\bar X=\bar X X=0$, where $\bar X^{AB}\equiv X_m\bar \G^{mAB}=\frac 1 2 \e^{ABCD}X_{CD}$.

Arbitrary 4d Lorentz representations can be built from products of spinors. So if we can lift spinor operators to the embedding space, we can lift any representation.  As shown in \cite{Weinberg:2010fx}, spinors lift to twistors.  Specifically, given a spinor primary $\psi_\a(x)$ with dimension $\De$, the combination
\begin{align}
\Psi_A(X) &\equiv (X^+)^{1/2-\De}\left(
\begin{array}{c}
\psi_\a(x)\\
i(x\.\bar\s)^{\dot\a\b}\psi_\b(x)
\end{array}
\right),
\label{eq:firstliftspinor}
\end{align}
with $x^\mu=X^\mu/X^+$, transforms as a twistor under the conformal group.  By construction, $\Psi_A(X)$ satisfies the transverseness condition $\bar X^{AB}\Psi_B(X)=0$, and has degree $1/2 - \De$ in $X$.

It is convenient to use a slightly different (but equivalent) lift of $\psi_\a(x)$.  Note that we can always solve the transverseness condition $\bar X\Psi=0$ as $\Psi=X\bar \Psi$ for  some $\bar \Psi\in\bar \bT$. In turn, $\bar\Psi$ is defined modulo twistors of the form $\bar XZ$, $Z\in \bT$. This follows because the multiplication maps $X:\bar\bT\to\bT$ and $\bar X:\bT\to\bar\bT$ have rank two and compose to zero, so that 
\begin{align}
\ker(\bar X)=\mathrm{im}(X)\cong\bar\bT/\ker(X)=\bar\bT/\mathrm{im}(\bar X).
\end{align}
Solving the transverseness equation for~(\ref{eq:firstliftspinor}) in this way, we lift $\psi_\a(x)$ to a gauge-redundant dual-twistor of degree $-1/2 - \De$,
\begin{align}
\psi_\a(x) &\to \bar\Psi^A(X),\quad \textrm{where}\quad\bar\Psi(X)\sim \bar\Psi(X)+\bar XZ.
\end{align}
Similarly, right-chiral spinors $\bar\l_{\dot\a}(x)$ lift to twistors $\Lambda_A(X)$ of degree $-1/2-\De$ with a gauge-redundancy $\L(X)\sim\L(X)+X\bar Z$.

The relation between the original four-dimensional fields and their twistor counterparts is extremely simple,
\begin{align}
\label{eq:spinorsbackto4d1}
\psi_\a(x) &= X_{\a B}\bar\Psi^B(X)|_{X=(1,x^2,x^\mu)}\\
\label{eq:spinorsbackto4d2}
\bar\l_{\dot\a}(x) &= \bar X_{\dot\a}{}^B\Lambda_{B}(X)|_{X=(1,x^2,x^\mu)},
\end{align}
where in each case we restrict $X$ to the Poincar\'e section.  As an example, a two-point function of twistor fields is fixed by conformal invariance and homogeneity to have the form
\begin{align}
\<\bar\Psi^A(X) \Lambda_B(Y)\> &= \frac{\de^A_B}{(-2X\.Y)^{\De+1/2}},
\end{align}
where the gauge-redundancies of $\bar\Psi$ and $\L$ let us discard terms proportional to $\bar X^{AC}Y_{CB}=-2X\.Y-\bar Y^{AC}X_{CB}$.  Applying the dictionary (\ref{eq:spinorsbackto4d1}, \ref{eq:spinorsbackto4d2}), we find 
\begin{align}
\<\psi_\a(x)\bar\l_{\dot\b}(y)\> &=
-\frac{X_{\a A} \bar Y^{A}{}_{\dot\b}}{(-2X\.Y)^{\De+1/2}}= -\frac{i(x-y)_{\a\dot\b}}{(x-y)^{2\De+1}},
\end{align}
which is precisely the correct form for a two-point function of spinor primaries.

More general operators $\cO_{\a_1\dots \a_{j}}^{\dot\b_1\dots \dot\b_{\bar\jmath}}(x)$ in $(j/2,\jbar/2)$ representations of the Lorentz group lift to symmetric multi-twistors $\cO_{B_1\dots B_{\bar\jmath}}^{A_1\dots A_{j}}(X)$ of degree $-\De-j/2-\jbar/2$ in $X$, subject to a gauge-redundancy in each index.  As in section~\ref{sec:spinningcorrelatorsformalism}, it will often be useful to adopt index-free notation
\begin{align}
\cO(X,S,\bar S) &\equiv \cO_{B_1\dots B_\jbar}^{A_1\dots A_j}(X) S_{A_1}\cdots S_{A_j} \bar S^{B_1}\cdots \bar S^{B_\jbar}
\end{align}
where $S$ and $\bar S$ are auxiliary twistors.  In this language, the gauge-redundancy of $\cO$ means that we can restrict $S,\bar S$ to be transverse
\begin{align}
\label{eq:transverseauxiliarytwistor}
\bar X S=0,\qquad X\bar S=0.
\end{align}
Of course, these conditions can always be solved as $S=X\bar T$, $\bar S=\bar X T$ for some $T, \bar T$. Consequently, the product $\bar S S$ vanishes as well.  Going back to explicit indices, this means that $\cO^{A_1\dots A_j}_{B_1\dots B_{\jbar}}$ must also have a gauge redundancy under shifts proportional to $\de^{A_i}_{B_{\bar\imath}}$.

Given a multi-twistor field $\cO(X,S,\bar S)$, one can project back to four-dimensions as follows,
\begin{align}
\label{eq:general4ddictionary}
\cO_{\a_1\dots\a_j}^{\dot\a_1\dots\dot\a_\jbar}(x) &\equiv& \frac{1}{j!\jbar!}\p{X\pdr{}{S}}_{\a_1}\cdots\p{X\pdr{}{S}}_{\a_j}\p{\bar X\pdr{}{\bar S}}^{\dot\a_1}\cdots\p{\bar X\pdr{}{\bar S}}^{\dot\a_\jbar}\cO(X,S,\bar S),
\end{align}
where we restrict $X$ to the Poincar\'e section.

As a special case, a vector operator $j^\mu(x)$ can be represented in the embedding space either as a multi-twistor $J_A^B(X)$, or as a vector $\cJ_m(X)$ satisfying the conditions of section~\ref{sec:spinningcorrelatorsformalism}.  The relation between these two formalisms is
\begin{align}
\label{eq:vectortospinors}
\bar X^{AC}J_C^B(X)-\bar X^{BC}J_C^A(X) &= -i\bar \G^{mAB}\cJ_m.
\end{align}
This is consistent with the gauge redundancies in both descriptions.  The transformation $J^B_A\to J^B_A+X_{AC}\Lambda^{CB}$ acts trivially on the left-hand side, while the redundancies $J^B_A\to J^B_A+\l \de^B_A$ and $J^B_A\to J^B_A+\bar X^{BC}\Lambda_{CA}$ become shifts $\cJ_m\to \cJ_m+ X_m$. 

\subsection{Two-Point and Three-Point Functions}

In this section, we identify the basic ingredients for two- and three-point correlators of multi-twistor operators $\cO(X,S,\bar S)$.
Given the condition (\ref{eq:transverseauxiliarytwistor}), only one type of conformal invariant other than $X_{ij}$ can appear in a two-point function,
\be
\label{eq:twopointtwistorinvariants}
I_{i\bar j} &\equiv S_i\bar S_j,\qquad i\neq j.
\ee
For example, as we saw above, a two-point function of spinors is given by
\begin{align}
\<\Psi(X_1,S_1)\bar\Psi(X_2,\bar S_2)\> &= \frac{I_{1\bar 2}}{X_{12}^{\De+1/2}}.
\end{align}
Similarly, a dimension-$\De$ operator $\cO(X,S,\bar S)$ transforming in a $(j/2,\jbar/2)$ representation of the Lorentz group has two-point function
\begin{align}
\label{eq:generaltwistortwoptfunction}
\<\cO(X_1,S_1,\bar S_1)\bar \cO(X_2,S_2,\bar S_2)\> &= \frac{I_{1\bar 2}^{j}I_{2\bar 1}^{\jbar} }{X_{12}^{\De+j/2+\jbar/2}},
\end{align}
where $\bar \cO$ transforms in the $(\bar\jmath/2,j/2)$ Lorentz representation.

More invariants are possible in three-point correlators:
\begin{align}
\label{eq:threepointtwistorinvariants}
J_{i,jk} &\equiv S_i \bar X_j X_k \bar S_i \\
K_{ijk} &\equiv S_i \bar X_j S_k\\
\bar K_{ijk} &\equiv \bar S_i X_j\bar S_k,
\end{align}
where each vanishes unless $i\neq j\neq k$.  $J_{i,jk}$ is antisymmetric in its last two indices, while $K_{ijk}$ and $\bar K_{ijk}$ are antisymmetric under the exchange $i\leftrightarrow k$.

General three-point functions can be constructed from the invariants $I,J,K,\bar K$, along with the $X_{ij}$.  However, these invariants are not algebraically independent.  For instance, one can verify the relations
\begin{align}
\label{eq:invariantrelation1}
K_{123}\bar K_{231} =&  I_{3\bar 2}J_{1,23}-X_{23}I_{3\bar 1}I_{1\bar 2}\\
J_{2,31}K_{123} =& I_{1\bar 2} K_{312} X_{23}-I_{3\bar 2} K_{231} X_{12} \\
J_{2,31}\bar K_{123} =& I_{2\bar 3} \bar K_{231} X_{12}- I_{2\bar 1} \bar K_{312} X_{23}\\
J_{1,23}J_{2,31}J_{3,12} =& X_{12}X_{23}X_{31}(I_{1\bar 2} I_{2\bar 3} I_{3\bar 1}-I_{1\bar 3} I_{2\bar 1} X_{3\bar 2})\nn\\
 & -I_{1\bar 3}I_{3\bar 1} J_{2,31}X_{12}X_{23} -I_{2\bar 3}I_{3\bar 2} J_{1,23}X_{12}X_{31} -I_{1\bar 2}I_{2\bar 1} J_{3,12}X_{23}X_{31},
 \label{eq:invariantrelation4}
\end{align}
and arbitrary permutations of the labels $\{1,2,3\}$.  Additional relations are possible (in addition to those generated by the above).  We will not attempt to classify them here.\footnote{
In verifying (\ref{eq:invariantrelation1}-\ref{eq:invariantrelation4}), it's extremely convenient to use twistor coordinates on the null-cone $X_{AB}=Z_{A}W_{B}-Z_{B}W_{A}$.  Auxiliary spinors $S,\bar S$ can then be written
\begin{align}
S_A &= \a Z_A+\b W_A\\
\bar S^A &= \e^{ABCD}Z_{B}W_{C}(\g T_{D}+\de U_{D})
\end{align} 
for constants $\a,\b,\g,\de\in \C$.  Here $T,U$ are any two linearly-independent twistors, defined modulo $Z,W$.  For instance in an $n$-point function, we are free to choose $T_i=Z_{i+1}, U_i=W_{i+1}$.  One can additionally use the $\GL(2,\C)$ redundancy rotating $Z$ into $W$ to set $\b$ to zero.  Relations between invariants then follow from the Schouten identity
\begin{align}
0 &= \<1234\>\<5|+\<2345\>\<1|+\<3451\>\<2|+\<4512\>\<3|+\<5123\>\<4|,
\end{align}
which expresses the fact that any five twistors are linearly dependent.
}

The general form of any correlator is determined by which combinations of gauge- and conformal-invariants have the correct homogeneity properties.  As an example, let us consider a symmetric tensor $J(X,S,\bar S)$ of spin-$\ell$ and dimension $\De$ and its correlators with scalars.  The two-point function $\<J(X_1,S_1,\bar S_1)J(X_2,S_2,\bar S_2)\>$ is given by~(\ref{eq:generaltwistortwoptfunction}) with $j=\jbar=\ell$.  Projecting to flat space with~(\ref{eq:general4ddictionary}), this becomes
\begin{align}
\label{eq:spinLtwopointfunctionflatspace}
\<J^{\a_1\dots\a_\ell}_{\dot\a_1\dots\dot\a_\ell}(x)J^{\b_1\dots\b_\ell}_{\dot\b_1\dots\dot\b_\ell}(0)\>
&=
\frac{x^{(\a_1}{}_{\dot\b_1}\cdots x^{\a_\ell)}{}_{\dot\b_\ell}x^{(\b_1}{}_{\dot\a_1}\cdots x^{\b_\ell)}{}_{\dot\a_\ell}}{x^{2(\De+\ell)}}.
\end{align}
This normalization differs from the one in \cite{DO1,DO2}, $J_\mathrm{ours}= (-2)^{-\ell/2} J_\mathrm{theirs}$.  As a consequence, our conformal block normalizations will differ as well.

The only structure with the correct homogeneity properties for a three-point function of $J$ with scalars $\f_1,\f_2$ is
\begin{align}
\<\f_1(X_1)\f_2(X_2)J(X_3,S_3,\bar S_3)\> &= \l \frac{J_{3,12}^\ell}{
X_{12}^{\frac{\De_1+\De_2-\De+\ell}{2}}
X_{23}^{\frac{\De+\De_2-\De_{1}+\ell}{2}}
X_{13}^{\frac{\De+\De_{1}-\De_2+\ell}{2}}
},
\end{align}
where $\l$ is an OPE coefficient.  Restricting to the Poincar\'e section, and applying (\ref{eq:general4ddictionary}), this takes the familiar form
\begin{align}
\<\f_1(x_1)\f_2(x_2)J^{\a_1\dots \a_\ell}_{\dot\a_1\dots\dot\a_\ell}(x_3)\> &= \l 
\frac{V_3^{(\a_1}{}_{\dot\a_1}\cdots V_3^{\a_\ell)}{}_{\dot\a_\ell}}{
x_{12}^{\De_1+\De_2-\De+\ell}x_{23}^{\De+\De_2-\De_1-\ell}x_{13}^{\De+\De_1-\De_2-\ell}
},
\\
\textrm{where}\quad V_{3\a\dot\a} &=-\frac{(X_3\bar X_1 X_2 \bar X_3)_{\a\dot\a}}{X_{13}X_{23}}=i\frac{(x_{31})_{\a\dot\a}}{x_{31}^2}-i\frac{(x_{32})_{\a\dot\a}}{x_{32}^2}.
\end{align}
With the normalization convention~(\ref{eq:spinLtwopointfunctionflatspace}), $J$ is imaginary when $\ell$ is odd, so that $\l$ is always real in a unitary theory.

\subsection{Twistor Projectors and Shadows}

Given a multi-twistor operator $\cO$ with dimension $\De$ and Lorentz representation $(j,\bar\jmath)$, there is an essentially unique gauge- and conformally-invariant projector,
\begin{align}
|\cO|
&=
\frac{1}{j!^2\jbar!^2}\int D^4 X D^4 Y |\cO(X,S,\bar S)\>\frac{(\lptl_SX\bar Y\rptl_{\bar T})^{j}(\lptl_{\bar S}\bar X Y \rptl_T)^{\bar\jmath}}{(-2X\.Y)^{4-\De+j/2+\bar \jmath/2}}\<\bar\cO(Y,T,\bar T)|\\
&= \frac{1}{j!^2\jbar!^2}\int D^4 X |\cO(X,S,\bar S)\>(\lptl_S X\rptl_T)^{j}(\lptl_{\bar S} \bar X\rptl_{\bar T})^{\bar\jmath}\<\tl\cO(X,T,\bar T)|,
\end{align}
The products $X\bar Y$ and $\bar X Y$ in the numerator are required to project away gauge-dependent pieces.  They are analogous to the factors $\eta^{mn}(X\.Y)-Y^m X^n$ in the tensor projector~(\ref{eq:spinprojector1}).  In the second line, we have defined the the shadow operator $\tl\cO$ by
\begin{align}
\tl\cO(X,S,\bar S) &\equiv \int D^4 Y \frac{1}{(-2X\.Y)^{4-\De+j/2+\bar \jmath/2}}\bar\cO(Y,Y\bar S,\bar Y S).
\end{align}
Note that $Y\bar S$ and $\bar Y S$ are automatically transverse with respect to $Y$, so that $\tl\cO$ is well-defined.  Formally, $\tl\cO$ has the properties of a primary operator of dimension $4-\De$ transforming in the $(\bar\jmath/2,j/2)$ representation of the Lorentz group.  As usual, this is useful in constraining the form of correlators involving $\tl\cO$.

\subsection{Example: Spin-$\ell$ Exchange between Scalars}

With the projector $|\cO|$ in hand, we can specialize the procedure in section~\ref{sec:higherspinblocksgeneralmethod} to compute conformal blocks of multi-twistor operators.  As a first example, let us reproduce the known conformal block for exchange of a spin-$\ell$ operator $J$ with dimension $\De$ between scalars $\f_i$.  We will assume that $J$ is normalized as in (\ref{eq:generaltwistortwoptfunction}).

Beginning with the three-point function
\begin{align}
\<J(X_0,S,\bar S)\f_3(X_3)\f_4(X_4)\>
=& \frac{(S\bar X_3 X_4 \bar S)^\ell}{
X_{34}^{\frac{\De_3+\De_4-\De+\ell}{2}}
X_{03}^{\frac{\De+\De_{34}+\ell}{2}}
X_{04}^{\frac{\De-\De_{34}+\ell}{2}}
},
\end{align}
one can compute
\begin{align}
\label{eq:spinLtwistorshadow}
\<\tl J(X_0,T,\bar T)\f_3(X_3)\f_4(X_4)\> =& (-1)^\ell \frac{\pi^2 \G(\De+\ell-1)}{\G(\tl\De+\ell)(\De-2)}
\frac{\G\big(\frac{\tl\De-\De_{34}+\ell}{2}\big)\G\big(\frac{\tl\De+\De_{34}+\ell}{2}\big)}{\G\big(\frac{\De-\De_{34}+\ell}{2}\big)\G\big(\frac{\De+\De_{34}+\ell}{2}\big)}\nn\\
&\x\frac{(T\bar X_3 X_4\bar T)^\ell}{
X_{34}^{\frac{\De_3+\De_4-\tl\De+\ell}{2}}
X_{03}^{\frac{\tl\De+\De_{34}+\ell}{2}}
X_{04}^{\frac{\tl\De-\De_{34}+\ell}{2}}
},
\end{align}
where $\tl\De=4-\De$.  As expected, this has the form of a three-point function between scalars and a spin-$\ell$ operator of dimension $\tl\De$.

Before sewing three-point correlators to compute a conformal block, we must determine the correct normalization factor.  (Since there is a unique allowed three-point structure, the mixing matrix $\cM$ is simply an overall constant.)
Inserting the projector $|J|$ between a two- and a three-point function is equivalent to iterating the shadow transform twice, so we can read off the correct normalization factor from~(\ref{eq:spinLtwistorshadow}),
\begin{align}
\cM\<J\f_3\f_4\>
&= \<J|J|\f_3\f_4\> \nn\\
&= \<\,\tl{\tl{\,\!\!J}}\f_3\f_4\>\nn\\
&= \frac{\pi^2\G(\De+\ell-1)}{\G(\tl\De+\ell)(\De-2)}\frac{\pi^2\G(\tl\De+\ell-1)}{\G(\De+\ell)(2-\De)}\<J\f_3\f_4\>
\end{align}

Finally, the numerator of our conformal block integral is given by
\begin{align}
\label{eq:gegenbauernumerator}
(S \bar X_1 X_2 \bar S)^\ell\frac{(\lptl_S X_0\rptl_T)^{j}(\lptl_{\bar S} \bar X_0\rptl_{\bar T})^{\bar\jmath}}{\ell!^4}(T\bar X_3 X_4\bar T)^\ell
&= (-1)^\ell s^{\ell/2}C^1_\ell(t),
\end{align}
where $C_\ell^\l(t)$ are Gegenbauer polynomials and
\begin{align}
t &\equiv
\frac{-X_{13}X_{02}X_{04}}{2\sqrt{s}}-(1\leftrightarrow 2)-(3\leftrightarrow 4),\\
s &\equiv X_{01}X_{02}X_{03}X_{04}X_{12}X_{34}.
\end{align}
Putting everything together, we have
\begin{align}
\p{\frac{X_{14}}{X_{13}}}^{\frac{\De_{34}}{2}}&\p{\frac{X_{24}}{X_{14}}}^{\frac{\De_{12}}{2}}
\frac{g^{\De_i}_{\De,\ell}(u,v)}{X_{12}^{\frac{\De_1+\De_2}{2}}X_{34}^{\frac{\De_3+\De_4}{2}}}
=\frac{1}{\cM}\<\f_1(X_1)\f_2(X_2)|J|\f_3(X_3)\f_4(X_4)\>\\
&= \frac{\G(\De+\ell)(2-\De)}{\pi^2\G(\tl \De+\ell-1)}
\frac{\G\big(\frac{\tl\De-\De_{34}+\ell}{2}\big)\G\big(\frac{\tl\De+\De_{34}+\ell}{2}\big)}{\G\big(\frac{\De-\De_{34}+\ell}{2}\big)\G\big(\frac{\De+\De_{34}+\ell}{2}\big)}X_{12}^{-\frac{\De_1+\De_2-\De}{2}}X_{34}^{-\frac{\De_3+\De_4-\tl\De}{2}}\nn\\
&\x
\left.\int D^4 X_0\,\frac{C_\ell^1(t)}{
X_{10}^{\frac{\De+\De_{12}}{2}}
X_{20}^{\frac{\De-\De_{12}}{2}}
X_{30}^{\frac{\tl\De+\De_{34}}{2}}
X_{40}^{\frac{\tl\De-\De_{34}}{2}}
}\right|_{M=1}.
\label{eq:biggegenbauerintegral}
\end{align}
Expanding the polynomial $C_\ell^1(t)$, the above integral becomes a sum of basic conformal four-point integrals (\ref{eq:basicfourpointintegral}) which can be expressed in terms of hypergeometric functions using~(\ref{eq:evendimresult}).  Luckily, the work of simplifying the resulting sum has already been performed in \cite{DO1}, using a recursion relation for Gegenbauer polynomials along with elementary hypergeometric function identities.  The result is~(\ref{eq:DOscalarblock}), which we reproduce here for the reader's convenience\footnote{Note that the $g_{\De,\ell}^{\De_i}$ quoted here differs by a factor of $2^\ell$ from the one derived in \cite{DO1}.  This is a consequence of our two-point function normalization (\ref{eq:spinLtwopointfunctionflatspace}).  We retain the factor of $(-1)^\ell$ because we have also chosen conventions where three-point function coefficients are real in unitary theories.}
\begin{align}
g_{\De,\ell}^{\De_i}(z,\bar z) &=(-1)^\ell \frac{z\bar z}{z-\bar z}(k_{\De+\ell}(z)k_{\De-\ell-2}(\bar z)-(z\leftrightarrow \bar z))\nn\\
k_\b(x) &\equiv x^{\b/2}{}_2F_1\p{\frac{\b-\De_{12}}{2},\frac{\b+\De_{34}}{2},\b,x}.
\end{align}

It is not obvious from this derivation why~(\ref{eq:biggegenbauerintegral}) should telescope into such a compact form.  We expect there should exist a simpler route to the correct answer, perhaps beginning by expressing the conformal integral over $X_0$ in twistor variables.  This is clearly unnecessary in the case of conformal blocks for external scalars, since there the conformal Casimir equation can be solved directly (bypassing the calculation given here).  However, it could prove helpful in simplifying expressions for higher spin conformal blocks.  We leave further investigation of this idea to future work.

\subsection{Example: Antisymmetric Tensor Exchange between Vectors}

As an example that brings together all of the machinery in this section, let us consider the exchange of a self-dual antisymmetric tensor $F(X,S)$ (and its anti-self-dual conjugate $\bar F(X,\bar S)$) in a four-point function of vectors $J_i(X_i,S_i,\bar S_i)$.  This computation could also be performed using the tensor formalism of section~\ref{sec:spinningcorrelatorsformalism}, where the embedding space $\e$-tensor enters the self-duality condition for $F$.  In twistor language, the three independent structures that can appear in the three-point function $\<FJ_1J_2\>$ are 
\begin{align}
\label{eq:antisymtensorthreeptstructures}
\<F(X_0,S_0)J_1(X_1,S_1,\bar S_1)J_2(X_2,S_2,\bar S_2)\> =&
\phantom{+\ }\l_0\frac{I_{0\bar 1}I_{0\bar 2}K_{102}}{
X_{01}^{\frac{\De+\De_{1}-\De_2+2}{2}}
X_{02}^{\frac{\De+\De_{2}-\De_1+2}{2}}
X_{12}^{\frac{\De_1+\De_2-\De}{2}}
}\nn
\\
&+\l_1\frac{I_{0\bar 1}I_{1\bar 2}K_{012}}{
X_{01}^{\frac{\De+\De_{1}-\De_2+2}{2}}
X_{02}^{\frac{\De+\De_{2}-\De_1}{2}}
X_{12}^{\frac{\De_1+\De_2-\De+2}{2}}
}\nn
\\
&
+\l_2\frac{I_{0\bar 2}I_{2\bar 1}K_{021}}{
X_{01}^{\frac{\De+\De_{1}-\De_2}{2}}
X_{02}^{\frac{\De+\De_{2}-\De_1+2}{2}}
X_{12}^{\frac{\De_1+\De_2-\De+2}{2}}
}\\
=& \sum_{i=0,1,2} \<F J_1J_2\>^{(i)}\l_i.
\end{align}
Similarly, we have
\begin{align}
\<\bar FJ_3 J_4\> &= \sum_{i=0,1,2} \<\bar F J_3 J_4\>^{(i)}\bar \l_i,
\end{align}
where the structures $\<\bar F J_3 J_4\>^{(i)}$ are obtained from $\<FJ_1 J_2\>^{(i)}$ by replacing $1,2\to 3,4$ and conjugating the spinor invariants $I_{i\bar j}\to I_{j\bar i}, K_{ijk}\to\bar K_{ijk}$.

The shadow transform of $\<\bar F J_3 J_4\>$ is given by
\begin{align}
\label{eq:vectorvectorantisymtensorshadow}
\<\tl{\bar F}J_3 J_4\> &= \<FJ_3 J_4\>^{(i)}|_{\De\to\tl\De}\cS_i{}^j\bar\l_j,
\end{align}
where the structures $\<FJ_3 J_4\>^{(i)}|_{\De\to\tl\De}$ are those appearing in (\ref{eq:antisymtensorthreeptstructures}) with the replacements $1,2\to 3,4$, and $\De\to\tl\De$, and the matrix $\cS$ has entries
\begin{align}
\label{eq:vectorvectorantisymtensorshadowmatrix}
\mathcal{S} &=
\frac{\pi^2\G(\De-1)}{\G(\tl\De+1)}
\left(
\begin{array}{ccc}
-A^{1,1}_{1,1} & 0 & 0\\ 
\frac{(\De-2+\De_{34})}{2}
A^{1,0}_{1,1}
& 0 & A^{1,0}_{0,1}\\
\frac{(2-\De+\De_{34})}{2}
A^{0,1}_{1,1}
& A^{0,1}_{1,0} & 0
\end{array}
\right)
\\
A^{m,n}_{l,k} &\equiv\frac{\G(m+\frac{\tl\De-\De_{34}}{2})\G(n+\frac{\tl\De+\De_{34}}{2})}{
\G(l+\frac{\De-\De_{34}}{2})\G(k+\frac{\De+\De_{34}}{2})
}.
\end{align}
In deriving (\ref{eq:vectorvectorantisymtensorshadow}) and (\ref{eq:vectorvectorantisymtensorshadowmatrix}), we have used the relation
\begin{align}
\bar K_{304}K_{034}K_{043} &= K_{043}I_{0\bar 4}I_{4\bar 3}X_{03}-K_{034}I_{0\bar 3}I_{3\bar 4}X_{04}-K_{304}I_{0\bar 3}I_{0\bar 4}X_{34}.
\end{align}

As before, we can compute the appropriate mixing matrix by iterating the shadow transform twice,
\begin{align}
\<\bar F J_3 J_4\>^{(j)}\cM_j{}^{i}\l_i
&=\<\bar F|F|J_3 J_4\>\nn\\
&=\<\tl{\tl{\bar F}} J_3 J_4\>\nn\\
&=\<\bar F J_3 J_4\>^{(k)}(\cS|_{\De\to\tl\De})_k{}^j \cS_j{}^i\l_i\nn\\
&=\frac{\pi^4}{\De(\De-1)(\De-3)(\De-4)}\<\bar F J_3 J_4\>^{(i)}\l_i.
\end{align}
Here, it turns out that $\cM_i{}^j$ is proportional to the identity matrix.  This is in fact a general result for the exchange of any operator in a completely left-handed $(j/2,0)$ or completely right-handed $(0,\jbar/2)$ representation of the Lorentz group.

The conformal partial wave for exchange of $F$ in a four-point function of currents $J_i$ is then given by
\begin{align}
&\l_i\bar \l_j
\p{\frac{X_{14}}{X_{24}}}^{\frac{\De_{12}}{2}}
\p{\frac{X_{13}}{X_{14}}}^{\frac{\De_{34}}{2}}
\frac{
g^{i,j}_F(X_i,S_i,\bar S_i)
}{
X_{12}^{\frac{\De_1+\De_2}{2}+1}
X_{34}^{\frac{\De_3+\De_4}{2}+1}
}\nn\\
&= \l_i{}^{(i)}\<J_1J_2|F|J_3J_4\>^{(k)}(\cM^{-1})_{k}{}^j\bar\l_j\\
&= \frac{\De(\De-1)(\De-3)(\De-4)}{\pi^4}\frac{1}{2!^2}\int D^4 X\<J_1 J_2 F(X,S)\>(\lptl_S X\rptl_T)^2\<\tl{\bar F}(X,T) J_3 J_4\>.
\end{align}
This is a sum of tensor four-point integrals of the form~(\ref{eq:basictensorfourptintegrals}), which can be evaluated using (\ref{eq:basictensorfourptresult}) and (\ref{eq:evendimresult}).  The full $3\x3$ matrix of conformal blocks $g^{i,j}_F$ contains approximately a hundred terms, so for brevity we will present only a single component in the main text,
\begin{align}
g^{1,1}_F =&\frac{u^\De \De}{(\De-1)(\De-2)(\De+\De_{34})}\left[\phantom{\frac {\frac 1 2}{\frac 1 2}}\right.\nn\\
&I_{1\bar 2}I_{2\bar 1}I_{3\bar 4}I_{4\bar 3}\left(
-F^{(4)}_{3,2;2}+\frac{(1+u-v)(\De-\De_{34})}{2}F^{(4)}_{2,2;1}-\frac{u(\De-\De_{34})(2+\De-\De_{34})}{4}F^{(4)}_{1,2;0}
\right)\nn\\
&-\frac{I_{1\bar 2}I_{4\bar 3}K_{213}\bar K_{1 3 4}}{X_{13}}F^{(4)}_{3,2;2}
+I_{1\bar 2}I_{4\bar 3}\p{\frac{K_{243}\bar K_{134}}{X_{34}}+\frac{K_{213}\bar K_{124}}{X_{12}}}\frac{u(\De-\De_{34})}{2}F^{(4)}_{2,2;1}\nn\\
&\left.-\frac{I_{1\bar 2}I_{4\bar 3}K_{243}\bar K_{124}}{X_{24}}\frac{u(\De-\De_{34})(2+\De-\De_{34})}{2}F^{(4)}_{1,2;0}\right],
\end{align}
where
\begin{align}
F^{(h)}_{m,n;k}(z,\bar z) &\equiv \frac{\p{\frac{\De-\De_{12}}{2}}_m\p{\frac{\De+\De_{34}}{2}}_n}{(\De)_k^2}\p{-\pdr{}{v}}^{h-1}k_{m,n;k}(z)k_{m,n;k}(\bar z)\\
k_{m,n;k}(x) &\equiv{}_2 F_1\p{\frac{\De-\De_{12}}{2}-m,\frac{\De+\De_{34}}{2}-n,\De-k,x}.
\end{align}

\section{Discussion}
\label{sec:conclusion}

The strategy for computing higher-spin conformal blocks is as follows:
\begin{enumerate}
\item Lift primary operators to the embedding space.
\item Find a gauge- and conformally-invariant projector $|\cO|$.
\item Determine the proper normalization (or mixing matrix) by inserting $|\cO|$ within a three-point function.  
\item Conformal blocks are then given by inserting $|\cO|$ within a four-point function.  Perform the monodromy-projected conformal integrals using the formulae in section~\ref{sec:conformalintegrals}.
\end{enumerate}
We have shown how to apply this strategy to tensor operators in $d$-dimensions, and arbitrary operators in 4d, where we introduced an efficient formalism for writing down conformally-invariant correlators using auxiliary twistors.  But it should apply equally well in any setting.  In particular, it would be interesting to apply it to superconformal theories, perhaps providing a way to bypass the complicated superconformal block calculations in \cite{Dolan:2001tt,Poland:2010wg,Fortin:2011nq}.  This would require generalizing the notion of conformal integrals to superconformal integrals.  For 4d CFTs, the underlying twistor structure of the projective null-cone, and the superembedding formalism of \cite{Goldberger:2011yp} may play an important role.  Efficient methods for computing superconformal blocks could be especially valuable in six dimensions, where the bootstrap might shed light on the mysterious $\cN=(2,0)$ M5 brane SCFT.

An important task for applying bootstrap methods to higher-spin operators is now to compute all conformal blocks that can appear in a given four-point function.  For example, an OPE of currents $J_1^\mu \x J_2^\nu$ in four dimensions can contain any operator transforming in a Lorentz representation of the form $(\frac{j}{2},\frac{j}{2}), (\frac{j+2}{2},\frac{j}{2}), (\frac{j}{2},\frac{j+2}{2}), (\frac{j+ 4}{2},\frac{j}{2}), (\frac{j}{2},\frac{j+ 4}{2})$, $j\geq 0$.  The $(\frac{j}{2},\frac{j}{2})$ operators are traceless symmetric tensors, and their conformal blocks can be derived easily using the methods of \cite{Costa:2011dw}.  We have given as an example the computation for $(0,1)$ and $(1,0)$ operators.  However, to apply bootstrap methods to a four-point function of currents, we need conformal blocks for {\it all} possible operators, so a formidable task is in store.  

Given the formulae in section~\ref{sec:conformalintegrals}, our methods are algorithmic and can be readily computerized.  However, there is also reason to believe that compact analytic expressions might exist even for very general classes of conformal blocks.  In particular, we have not shed light on why the terms in the conformal block for spin-$\ell$ exchange between external scalars can be combined into such a simple form~(\ref{eq:DOscalarblock}).  Dolan and Osborn understood this using the conformal Casimir equation.  Although their argument becomes intractable in the case of higher spin, it's likely that similar structure is present.

Finally, let us note that the technology developed here should work equally well in Mellin space \cite{Mack:2009mi,Penedones:2010ue}, which is proving to be a convenient setting for understanding effective CFTs \cite{Heemskerk:2009pn,Fitzpatrick:2010zm} dual to weakly coupled theories in AdS \cite{Fitzpatrick:2011ia,Paulos:2011ie,Fitzpatrick:2011hu,Fitzpatrick:2011dm}.  The only modification would be the expressions for conformal integrals~(\ref{eq:intermediateintegral1}, \ref{eq:evendimresult}), which become functions of Mellin variables $\de_{ij}$ instead of conformal cross-ratios.  Higher spin conformal blocks in Mellin space could be useful for understanding the gauge and gravity sectors of effective CFTs.

\section*{Acknowledgements}

I am grateful to J. Bourjaily, C. C\'ordova, R. Loganayagam, D. Poland, S. Raju, and S. Rychkov for discussions and comments.  Thanks to Miguel Costa and Tobias Hansen for checking several calculations and pointing out errors in a previous version of this paper. This work is supported by NSF grant PHY-0855591 and the Harvard Center for
the Fundamental Laws of Nature.

\appendix

\section{Monodromy Projections and the OPE}
\label{app:monodromyandOPE}

The prescription~(\ref{eq:conformalblockfinalexpression}) for ensuring the conformal block's consistency with the OPE may seem somewhat ad hoc, so let us clarify why it is needed.  Along the way, we will elucidate the origin of the shadow contribution $g_{\tl\cO}$.  Recall that our ``candidate" conformal block for the exchange of a dimension-$\De$ scalar $\cO$ in a four-point function of dimension-$\de$ scalars $\f$ is given by
\begin{align}
F(X_i) &\propto X_{12}^{\de}X_{34}^\de \int D^d X D^d Y\<\f(X_1)\f(X_2)\cO(X)\>\frac{1}{(-2X\.Y)^{d-\De}}\<\cO(Y)\f(X_3)\f(X_4)\>.
\label{eq:integralforcandidate}
\end{align}

Why should~(\ref{eq:integralforcandidate}) violate the OPE in the first place?  This is clear already in the integrand: the objects $\<\f\f\cO\>$ are radially-ordered expectation values of fields.  In any given quantization, they include pieces where the $\f\x\f$ OPE is valid, and pieces where it is invalid.  For concreteness, restrict to the Poincar\'e section and consider radial quantization around the origin.  We may write
\begin{align}
\<\f(x_1)\f(x_2)\cO(x)\> &= 
\<0|\cR\{\f(x_1)\f(x_2)\cO(x)\}|0\>\\
&=\th(|x|>|x_1|,|x_2|)\<0|\cO(x)\cR\{\f(x_1)\f(x_2)\}|0\>\nn\\
&\phantom{=}+\th(|x|<|x_1|,|x_2|)\<0|\cR\{\f(x_1)\f(x_2)\}\cO(x)|0\>\nn\\
&\phantom{=}+\textrm{other orderings},
\label{eq:radiallyorderedthreeptfunction}
\end{align}
where $\cR\{\dots\}$ indicates radial-ordering.  In the first term, the $\f\x\f$ OPE is valid, since $\cO(x)$ lies outside a sphere surrounding $x_1,x_2$.  However, the $\f\x\f$ OPE does not converge in the other terms.\footnote{Of course, for a three-point function we can restore validity of the OPE by quantizing around a different point.  However, no single point ensures validity of the OPE for all values of $x_1,x_2$, and $x$.}

The different orderings in~(\ref{eq:radiallyorderedthreeptfunction}) are distinguished by their monodromy properties.  Specifically, consider the transformation $M=e^{2\pi i (\cD_1+\cD_2)}$, where $\cD=x\.\pdr{}{x}$ is the differential operator generating dilatations and $\cD_i$ indicates $\cD$ acting on the point $x_i$.  Clearly, $M x_{12}^2=e^{4\pi i}x_{12}^2$, while $M x_{ij}^2=x_{ij}^2$ for all other pairs $i,j$, assuming $x_3$ and $x_4$ are far from the origin.  If $\f(x)$ is primary with dimension $\de$, we have
\begin{align}
\label{eq:primarydilatationaction}
e^{\l(\cD+\de)}\f(x) &= e^{\l D}\f(x)e^{-\l D},
\end{align}
where $D$ generates dilatations on the Hilbert space.
Notice also that states $\cO(x)|0\>$ have energies of the form $\De+n$, where $n\in \Z$ (the primary state $|\cO\>$ has energy $\De$, while descendants $P^{\mu_1}\cdots P^{\mu_n}|\cO\>$ have energy $\De+n$).  Consequently, 
\begin{align}
\label{eq:monodromyoperatorphase1}
e^{\pm 2\pi i D}\cO(x)|0\> &= \cO(x)|0\>e^{\pm 2\pi i \De}\\
\label{eq:monodromyoperatorphase2}
\<0|\cO(x)e^{\pm 2\pi i D} &= e^{\pm 2\pi i \De}\<0|\cO(x).
\end{align}
Applying $M$ to the radially-ordered correlator and using these facts, each ordering picks up a different phase
\begin{align}
e^{2\pi i(\cD_1+\cD_2)}\<\f(x_1)\f(x_2)\cO(x)\> &= \th(|x|>|x_1|,|x_2|)\<0|\cO(x)\cR\{\f(x_1)\f(x_2)\}|0\>e^{2\pi i(\De-2\de)}\nn\\
&\phantom{=}+\th(|x|<|x_1|,|x_2|)\<0|\cR\{\f(x_1)\f(x_2)\}\cO(x)|0\>e^{2\pi i(-\De-2\de)}\nn\\
&\phantom{=}+\textrm{other orderings}.
\end{align}

We see that the $\<0|\cO\f\f|0\>$ ordering, where the OPE is valid, contributes precisely to the part of $F(X_i)$ with monodromy $e^{2\pi i\De}$, namely $g_{\cO}$. (When acting on $F(X_i)$ the phases $e^{2\pi i(-2\de)}$ are cancelled by the factor $x_{12}^{2\de}$ out front.)  Meanwhile, the $\<0|\f\f\cO|0\>$ ordering contributes to the shadow block $g_{\tl\cO}$ (assuming $d\in\Z$).  Thus, projection onto the correct monodromy eigenspace is equivalent to including the $\theta$-functions $\th(|x|>|x_1|,|x_2|)$ in the integral~(\ref{eq:opshadowopintegral}), carving out a sphere around $x_1,x_2$ and ensuring validity of the OPE.

The appearance of $\th$-functions in the integrand raises a puzzle. The form of these $\th$-functions depends on our choice of dilatation operator $\cD$, since different choices imply different radial orderings.  Thus, they na\"ively break conformal invariance.  However, the monodromy argument makes it clear that this breaking is somehow weak.  Monodromy projection introduces similar $\th$-functions in the conformal block (but they take the value 1 when $|x_{1,2}|\ll |x_{3,4}|$ so we have ignored them in the main text).  Somehow, changing the $\th$-functions in the integrand changes {\it only} these $\th$-functions in the result.  It would be interesting to understand why in more detail.


\section{Spinor Conventions in Six Dimensions}
\label{app:spinorconventions}

We choose $-+++$ signature for the metric $g_{\mu\nu}$ in 4d Minkowski space, and follow the conventions of Wess and Bagger \cite{Wess:1992cp} for four-dimensional spinors.  The six-dimensional embedding space metric is given by $\eta_{mn}X^m X^n = -X^+X^-+g_{\mu\nu}X^\mu X^\nu$.

Six-dimensional spinors (twistors) decompose under the 4d Lorentz group as
\begin{align}
Z_A &= \left(
\begin{array}{c}
\l_\a\\
\mu^{\dot\a}
\end{array}
\right).
\end{align}
We choose conventions where the $\SU(2,2)$-invariant antisymmetric tensors $\e_{ABCD}, \e^{ABCD}$ satisfy $\e_{1234}=\e^{1234}=+1$.  We have the antisymmetric chiral Gamma matrices
\begin{align}
\G^+_{AB}=\twobytwo{0}{0}{0}{2i\e^{\dot\a\dot\b}},
\quad
\G^-_{AB}=\twobytwo{-2i\e_{\a\b}}{0}{0}{0},
\quad
\G^\mu_{AB} = \twobytwo{0}{\s^\mu_{\a\dot\g}\e^{\dot\g\dot\b}}{-\bar\s^{\mu\dot\a\g}\e_{\g\b}}{0}.
\end{align}
And also $\tl\G^{mAB}=\frac 1 2 \e^{ABCD}\G^{m}_{CD}$, which are given by
\begin{align}
\tl\G^{+AB}=\twobytwo{2i\e^{\a\b}}{0}{0}{0},
\quad
\tl\G^{-AB}=\twobytwo{0}{0}{0}{-2i\e_{\dot\a\dot\b}},
\quad
\tl\G^{\mu AB} =\twobytwo{0}{-\e^{\a\g}\s^\mu_{\g\dot\b}}{\e_{\dot\a\dot\g}\bar\s^{\mu\dot\g\b}}{0}.
\end{align}
They satisfy
\begin{align}
(\G^m\tl\G^n+\G^n\tl\G^m)_A{}^B&= -2\eta^{nm}\de_A{}^B\\
(\tl\G^m\G^n+\tl\G^n\G^m)^A{}_B&= -2\eta^{nm}\de^A{}_B\\
\tl\G^{mAB}\G_{mCD}&= 2(\de^A_C\de^B_D-\de^B_C\de^A_D)\\
\tl \G^{mAB}\tl \G_m{}^{CD}&= 2\e^{ABCD}\\
\G^m_{AB}\G_{mCD}&= 2\e_{ABCD}.
\end{align}

With them, we can define bi-spinors
\begin{align}
\bar X^{AB}=X_m\tl\G^{mAB},\quad X_{AB}=X_m \G^m_{AB}.
\end{align}
The inverse transformation is
\begin{align}
X^m=\frac 1 4 X_{AB}\tl\G^{mAB}=\frac 1 4 \bar X^{AB}\G^m_{AB}.
\end{align}
We also have the inner product
\begin{align}
X^m Y_m = -\frac 1 4\Tr(\bar XY).
\end{align}


\bibliography{Biblio}{}
\bibliographystyle{utphys}

\end{document}